# A novel circular semiquantum private comparison protocol of equality without a pre-shared key based on χ-type states

Jiang-Yuan Lian, Tian-Yu Ye*

College of Information & Electronic Engineering, Zhejiang Gongshang University, Hangzhou 310018, P.R. China

E-mail: yetianyu@zjgsu.edu.cn (T.Y. Ye)

**Abstract:** In this paper, we adopt χ-type states to design a novel circular semiquantum private comparison (SQPC) protocol which can determine the equality of private inputs from two semiquantum users within one round implementation under the help of a semi-honest third party (TP) who possesses complete quantum capabilities. Here, it is assumed that the semi-honest TP has the abilities to launch all possible attacks to steal useful information about two semiquantum users' private inputs but cannot conspire with anyone else. The travelling particles go from TP to Alice, Alice to Bob and back from Bob to TP. The security analysis turns out the proposed SQPC protocol can resist both the outside attacks and the inside attacks. The proposed SQPC protocol has no demand for unitary operations. Compared with some existing SQPC protocols of equality with quantum entangled states, the proposed SQPC protocol has some advantages more or less: ① it requires no pre-shared key among different participants; ② it doesn't need quantum entanglement swapping; and ③ it employs no delay lines.

## 1 Introduction

Quantum information science, as one of the hottest research areas, has attracted plenty of attention of scholars during recent years. As a matter of fact, in the year of 1984, Bennett and Brassard [1] put forward the first quantum key distribution (QKD) protocol, commonly named as the BB84 protocol hereafter, by employing the polarization of single photons, which can be seen as a genuine beginning of quantum cryptography. Quickly, quantum cryptography has gained many branches deserving to research, such as quantum key distribution (QKD) [1-5], quantum secret sharing (QSS) [6-9], quantum private comparison (QPC) [10-17], and so on. Speaking of QPC, in the year of 2009, Yang and Wen [10] proposed the first QPC protocol which can implement the comparison of equality of private inputs from two different users by absorbing quantum mechanics into classical private comparison. Since then, a range of QPC protocols [11-17] have been proposed in turn with different quantum states. However, there is a common problem in these QPC protocols that all users are required to own complete quantum capabilities, which makes the users with limited quantum abilities unable to participant in the designed protocols.

In the year of 2007, the novel concept of semiquantumness was put forward by Boyer *et al.* [18,19]. As illustrated in Refs.[18-21], the semiquantum user who has limited quantum capabilities is restricted to execute the following operations: (1) sending or reflecting qubits; (2) measuring qubits in the $Z$ basis (i.e., $\{|0\rangle,|1\rangle\}$); (3) preparing qubits in the $Z$ basis; and (4) reordering the qubits through different delay lines. After the concept of semiquantumness was introduced into QPC, a series of semiquantum private comparison (SQPC) protocols [22-34] were proposed successively. In the year of 2016, Chou *et al.* [22] proposed the first SQPC protocol based on entanglement swapping of Bell states, which can determine the equality of private inputs from two semiquantum participants. In the year of 2018, by adopting Bell states, Thapliyal *et al.* [23] put forward a SQPC protocol of equality which requires a pre-shared key; and by using two-particle product states, Ye *et al.* [24] proposed a measure-resend SQPC protocol of equality which requires a pre-shared key. In the year of 2019, Lin *et al.* [25] utilized single photons to design a SQPC protocol of equality without a pre-shared key; and by adopting Bell states, Yan *et al.* [26] put forward a randomization-based SQPC protocol of equality with a pre-shared key. In the year of 2020, by adopting Bell states, two SQPC protocols, each of which needs a pre-shared key, were proposed by Jiang [27], where the first protocol needs the semiquantum users to measure the received particles but the second protocol has no this demand. In the year of 2021, Tsai *et al.* [28] and Xie *et al.* [29] pointed out that the first SQPC protocol in Ref.[27] has security loopholes and put forward the improved protocols accordingly, each of which requires a pre-shared key; by utilizing three-particle GHZ-like states; Ye *et al.* [30] designed an efficient SQPC protocol of equality without using entanglement resource and pre-shared key; and by employing Bell states, Sun *et al.* [31] proposed a novel measure-resend SQPC scheme of equality with a pre-shared key. In the year of 2022, Tian *et al.* [32] put forward a W-state-based SQPC protocol of equality with a pre-shared key; by employing single kind of Bell states, Geng *et al.* [33] put forward a novel SQPC protocol of equality with a pre-shared key. Apparently, the generation of a pre-shared key in a quantum communication protocol consumes extra quantum resources and classical resources. Hence, it is worthy of studying how to design a SQPC protocol without a pre-shared key.

In accordance with foregoing analysis, we try to utilize the $\chi$-type states as quantum resource to design a novel circular SQPC protocol of equality without a pre-shared key. In the proposed SQPC protocol, the third party (TP) is assumed to be semi-honest, which means that she can misbehave on her own but is not allowed to conspire with anyone else [12]. TP is required to own complete quantum abilities, while the two users owning private inputs are restricted to possess limited quantum capabilities. The travelling particles go from TP to Alice, Alice to Bob and back from Bob to TP. The proposed SQPC protocol has no requirement for unitary operations. Compared with the existing SQPC protocols of equality with quantum entangled states in Refs.[22,23,26-29,31-33], the proposed SQPC protocol has some advantages more or less: ① it requires no pre-shared key among different participants; ② it doesn't need quantum entanglement swapping; and ③ it employs no delay lines.

## 2 Protocol description

In a Hilbert space, the four Bell states $|\phi^{\pm}\rangle$ and $|\psi^{\pm}\rangle$ form a complete orthogonal basis, where

$$|\phi^{+}\rangle=\frac{1}{\sqrt{2}}\left(|00\rangle+|11\rangle\right), \tag{1}$$

$$|\phi^{-}\rangle=\frac{1}{\sqrt{2}}\left(|00\rangle-|11\rangle\right), \tag{2}$$

$$|\psi^{+}\rangle=\frac{1}{\sqrt{2}}\left(|01\rangle+|10\rangle\right), \tag{3}$$

and

$$|\psi^{-}\rangle=\frac{1}{\sqrt{2}}\left(|01\rangle-|10\rangle\right). \tag{4}$$

As depicted in Ref. [34], the χ-type state is defined as

$$|\chi^{00}\rangle_{1234}=\frac{1}{2\sqrt{2}}\left(|0000\rangle+|0011\rangle+|1100\rangle-|1111\rangle-|0101\rangle+|0110\rangle+|1001\rangle+|1010\rangle\right)_{1234} \tag{5}$$

$$=\frac{1}{2}\left(|00\rangle|\phi^{+}\rangle+|11\rangle|\phi^{-}\rangle-|01\rangle|\psi^{-}\rangle+|10\rangle|\psi^{+}\rangle\right)_{1234} \tag{6}$$

$$=\frac{1}{2}\left(|\phi^{+}\rangle|00\rangle+|\phi^{-}\rangle|11\rangle-|\psi^{-}\rangle|01\rangle+|\psi^{+}\rangle|10\rangle\right)_{1234}, \tag{7}$$

where subscripts 1, 2, 3 and 4 represent the four particles of a χ-type state in order. An orthonormal basis set in the four-qubit Hilbert space can be obtained by performing Pauli operations on particles 1 and 3 in $|\chi^{00}\rangle_{1234}$:

$$FMB=\left\{|\chi^{yz}\rangle_{1234}=\sigma_1^{y}\sigma_3^{z}|\chi^{00}\rangle_{1234}\,|y,z=0,1,2,3\right\}, \tag{8}$$

where $\sigma^0=|0\rangle\langle 0|+|1\rangle\langle 1|$, $\sigma^1=|0\rangle\langle 1|+|1\rangle\langle 0|$, $\sigma^2=|0\rangle\langle 1|-|1\rangle\langle 0|$ and $\sigma^3=|0\rangle\langle 0|-|1\rangle\langle 1|$ are four Pauli operations.

Assume that there are two semiquantum users with limited quantum abilities, Alice and Bob, who have the private input sequences $p_a=\{p_a^1,p_a^2,\ldots,p_a^L\}$ and $p_b=\{p_b^1,p_b^2,\ldots,p_b^L\}$, respectively, where $p_a^i,p_b^i\in\{0,1\}$ and $i=1,2,\ldots,L$. Alice and Bob want to compare the equality of their private input sequences with the help of a semi-honest TP who has complete quantum capacities. In the following, we will introduce the specific procedures of the proposed SQPC protocol in detail, which is also shown in Fig.1 for clarity after the security check processes are ignored. Note that in this protocol, $|0\rangle$ corresponds to the classical bit 0, while $|1\rangle$ corresponds to the classical bit 1.

Step 1: TP prepares $12L$ $\chi$-type states according to Eq.(5) and divides them into three different particle sequences $S_1$, $S_2$ and $S_3$. Specifically speaking, $S_1 = \{S_1^1, S_1^2, \ldots, S_1^{12L}\}$ and $S_2 = \{S_2^1, S_2^2, \ldots, S_2^{12L}\}$ are composed by all first particles and all second particles of these $12L$ $\chi$-type states, respectively. Here, $S_1^l$ is the first particle of the $l$ th $\chi$-type state while $S_2^l$ is the second particle of the $l$ th $\chi$-type state, where $l = 1, 2, \ldots, 12L$. And $S_3 = \{S_3^1, S_3^2, \ldots, S_3^{12L}\}$ are formed by the third and the fourth particles of these $12L$ $\chi$-type states, where $S_3^l$ represents the particle pair composed by the third and the fourth particles of the $l$ th $\chi$-type state. Then, TP selects the first $8L$ particles in $S_1$, transmits these selected particles to Alice one by one via the quantum channel and keeps the remaining particles in her hand. Note that TP sends out the next particle only after receiving the previous one.

Step 2: Alice randomly chooses the REFLECT_A mode or the MEASURE_A mode for each received particle from TP. The REFLECT_A mode means to reflect the received particles to Bob without any interference, while the MEASURE_A mode implies to measure the received particles with the $Z$ basis, record the measurement results, prepare the fresh quantum states according to the found states and send them to Bob. The sequence composed by the first $8L$ particles in $S_1$ after Alice's operations is represented by $S_1^{'}$.

Bob randomly chooses the REFLECT_B mode or the MEASURE_B mode for each received particle from Alice. Here, the REFLECT_B mode means to reflect the received particles back to TP without any interference, while the MEASURE_B mode implies to measure the received particles with the $Z$ basis, record the measurement results, prepare the fresh quantum states as found and send them back to TP. $S_1^{'}$ after Bob's operations is represented by $S_1^{''}$.

Step 3: TP receives all particles from Bob, stores them in a quantum memory and sends a confirm signal to Alice and Bob. Then Alice announces the positions where she chose the MEASURE_A mode, while Bob announces the positions where he chose the MEASURE_B mode. Alice, Bob and TP perform the corresponding actions according to the four different Cases in Table 1.

Case 1: Alice has entered into the REFLECT_A mode, while Bob has entered into the REFLECT_B mode. TP performs the $FMB$ basis measurements on the particles of $S_1^{''}$ corresponding to this Case, the corresponding particles of $S_2$ and the corresponding particle pairs of $S_3$. If there are no eavesdropping, TP's *FMB* basis measurement results should always be $\left|\chi^{00}\right\rangle$. TP calculates the error rate by comparing her measurement results in this Case with $\left|\chi^{00}\right\rangle$. If the error rate is higher than the threshold, the protocol will be aborted; otherwise, the protocol will be proceeded;

Case 2: Alice has entered into the MEASURE_A mode, while Bob has entered into the MEASURE_B mode. TP performs Bell basis measurements on the particle pairs of $S_3$ corresponding to this Case and $Z$ basis measurements on the particles of $S_2$ corresponding to this Case. Alice announces TP her measurement results on the particles of $S_1$ corresponding to this Case. In accordance with Eq.(6), TP calculates the error rate by judging whether her Bell basis measurement results on the particle pairs of $S_3$ corresponding to this Case and her $Z$ basis measurement results on the particles of $S_2$ corresponding to this Case are correctly correlated to Alice's measurement results on the particles of $S_1$ corresponding to this Case or not. If the error rate is higher than the threshold, the protocol will be aborted; otherwise, the protocol will be proceeded.

TP performs $Z$ basis measurements on the particles of $S_1^{''}$ corresponding to this Case. Bob announces TP his measurement results on the particles of $S_1^{'}$ corresponding to this Case. TP calculates the error rate by judging whether Alice's measurement results on the particles of $S_1$ corresponding to this Case, Bob's measurement results on the corresponding particles of $S_1^{'}$ and her $Z$ basis measurements on the corresponding particles of $S_1^{''}$ are same or not. If the error rate is higher than the threshold, the protocol will be aborted; otherwise, the protocol will be proceeded.

Case 3: Alice has entered into the REFLECT_A mode, while Bob has entered into the MEASURE_B mode. TP performs $Z$ basis measurements on the particles of $S_1^{''}$ corresponding to this Case. Bob randomly chooses $L$ measurement results on the particles of $S_1^{'}$ corresponding to this Case and announces TP their positions and values. TP performs Bell basis measurement on the corresponding particle pairs of $S_3$ and $Z$ basis measurements on the corresponding particles of $S_2$. According to Eq.(6), TP calculates the error rate by judging whether her $Z$ basis measurements on the corresponding particles of $S_1^{''}$, her Bell basis measurement results on the corresponding particle pairs of $S_3$ and her $Z$ basis measurement results on the corresponding particles of $S_2$ are correctly correlated to Bob's announced measurement results or not. If the error rate is higher than the threshold, the protocol will be aborted; otherwise, the protocol will be proceeded.

Bob's remaining $L$ measurement results on the particles of $S_1^{'}$ corresponding to this Case can be denoted as $m_b=\left\{m_b^1,m_b^2,\ldots,m_b^L\right\}$, where $m_b^i\in\{0,1\}$ and $i=1,2,\ldots,L$. Note that TP can easily deduce $m_b$ from her $Z$ basis measurements on the corresponding particles of $S_1^{''}$.

Case 4: Alice has entered into the MEASURE_A mode, while Bob has entered into the REFLECT_B mode. TP performs $Z$ basis measurements on the particles of $S_1^{''}$ corresponding to this Case. Alice randomly chooses $L$ measurement results on the particles of $S_1$ corresponding to this Case and announces TP their positions and values. TP performs Bell basis measurement on the corresponding particle pairs of $S_3$ and $Z$ basis measurements on the corresponding particles of $S_2$. TP calculates the error rate by judging whether her $Z$ basis measurements on the corresponding particles of $S_1^{"}$, her Bell basis measurement results on the corresponding particle pairs of $S_3$ and her $Z$ basis measurement results on the corresponding particles of $S_2$ are correctly correlated to Alice's announced measurement results or not. If the error rate is higher than the threshold, the protocol will be aborted; otherwise, the protocol will be proceeded.

Alice's remaining $L$ measurement results on the particles of $S_1$ corresponding to this Case can be denoted as $m_a = \{m_a^1, m_a^2, \ldots, m_a^L\}$, where $m_a^i \in \{0,1\}$ and $i = 1, 2, \ldots, L$. Note that TP can easily deduce $m_a$ from her $Z$ basis measurements on the corresponding particles of $S_1^{"}$.

Table 1 The usages of particles from $S_1$ under different Cases

| Case | The mode of Alice | The mode of Bob | Usage |
|---|---|---|---|
| Case 1 | The REFLECT_A mode | The REFLECT_B mode | Eavesdropping detection |
| Case 2 | The MEASURE_A mode | The MEASURE_B mode | Eavesdropping detection |
| Case 3 | The REFLECT_A mode | The MEASURE_B mode | Eavesdropping detection and private comparison |
| Case 4 | The MEASURE_A mode | The REFLECT_B mode | Eavesdropping detection and private comparison |

Step 4: After finishing sending out the first $8L$ particles in $S_1$, TP needs to send a signal to Alice and Bob, which means that she will transmit the last $4L$ particle pairs in $S_3$ to Alice one by one. Alice randomly selects the REFLECT_A mode or the MEASURE_A2 mode for each received particle pair from TP. The MEASURE_A2 mode implies to measure the received particles with the $Z$ basis, record the measurement results, prepare the fresh quantum states according to the found states or the bits of $M_A$ with equal probabilities and send them to Bob. Here, $M_A = \{M_A^1, M_A^2, \ldots, M_A^{2L}\}$ is a random binary sequence generated by Alice beforehand utilizing a random number generator, where $M_A^j \in \{0,1\}$, $j = 1, 2, \ldots, 2L$; when $M_A^j = 0$, the prepared fresh quantum state is $|0\rangle$, and when $M_A^j = 1$, the prepared fresh quantum state is $|1\rangle$. The sequence composed by the last $4L$ particle pairs in $S_3$ after Alice's operations is represented by $S_3^{'}$.

Bob randomly chooses the REFLECT_B mode or the MEASURE_B2 mode for each received particle pair from Alice. The MEASURE_B2 mode implies to measure the received particles with the $Z$ basis, record the measurement results, prepare the fresh quantum states in accordance with the bits of $M_B$ and send them back to TP. Here, $M_B = \{M_B^1, M_B^2, \ldots, M_B^{4L}\}$ is a random binary sequence generated by Bob beforehand utilizing a random number generator, where $M_B^k \in \{0,1\}$ , $k = 1, 2, \ldots, 4L$ ; when $M_B^k = 0$ , the prepared fresh quantum state is $|0\rangle$ , and when $M_B^k = 1$ , the prepared fresh quantum state is $|1\rangle$ . $S_3^{'}$ after Bob's operations is represented by $S_3^{''}$ .

Step 5: TP receives all particles from Bob in quantum memory, and sends a confirm signal to Alice and Bob. Then Alice announces the positions where she chose the MEASURE_A2 mode, while Bob announces the positions where he chose the MEASURE_B2 mode. In addition, Alice needs to publish the positions where she generated the fresh quantum states as found in $S_3$ . Alice, Bob and TP perform the corresponding actions according to the four different Cases in Table 2.

Case ①: Alice has entered into the REFLECT_A mode, while Bob has entered into the REFLECT_B mode. TP performs the *FMB* basis measurements on the particle pairs of $S_3^{''}$ corresponding to this Case, the corresponding particles of $S_1$ and the corresponding particles of $S_2$ . If there are no eavesdropping, TP's *FMB* basis measurement results should always be $\left|\chi^{00}\right\rangle$ . TP calculates the error rate by comparing her measurement results in this Case with $\left|\chi^{00}\right\rangle$ . If the error rate is higher than the threshold, the protocol will be aborted; otherwise, the protocol will be proceeded;

Case ②: Alice has entered into the REFLECT_A mode, while Bob has entered into the MEASURE_B2 mode. Bob announces TP his measurement results on $L$ particle pairs in $S_3^{'}$ corresponding to this Case. TP performs Bell basis measurements on the particles of $S_1$ corresponding to this Case and the particles of $S_2$ corresponding to this Case. Based on Eq.(7), TP calculates the error rate by judging whether her Bell basis measurement results on the corresponding particles of $S_1$ and $S_2$ are correctly correlated to Bob's announced measurement results or not. If the error rate is higher than the threshold, the protocol will be aborted; otherwise, the protocol will be proceeded.

After that, TP measures the particle pairs of $S_3^{''}$ corresponding to this Case with the $Z$ basis. Bob announces TP the corresponding $2L$ bits from $M_B$ . TP calculates the error rate by judging whether her $Z$ basis measurement results on the corresponding particle pairs of $S_3^{''}$ are correctly correlated to the corresponding $2L$ bits from $M_B$ or not. If the error rate is higher than the threshold, the protocol will be aborted; otherwise, the protocol will be proceeded.

Case ③: Alice has entered into the MEASURE_A2 mode, while Bob has entered into the REFLECT_B mode. Alice announces TP her measurement results on $L$ particle pairs in $S_3$ corresponding to this Case. TP performs Bell basis measurements on the particles of $S_1$ corresponding to this Case and the particles of $S_2$ corresponding to this Case. In the light of Eq.(7), TP calculates the error rate by judging whether her Bell basis measurement results on the corresponding particles of $S_1$ and $S_2$ are correctly correlated to Alice's announced measurement results or not. If the error rate is higher than the threshold, the protocol will be aborted; otherwise, the protocol will be proceeded.

Case ④: Alice has entered into the MEASURE_A2 mode, while Bob has entered into the MEASURE_B2 mode. Alice announces TP her measurement results on $L$ particle pairs in $S_3$ corresponding to this Case. TP performs Bell basis measurements on the particles of $S_1$ corresponding to this Case and the particles of $S_2$ corresponding to this Case. In accordance with Eq.(7), TP calculates the error rate by judging whether her Bell basis measurement results on the corresponding particles of $S_1$ and $S_2$ are correctly correlated to Alice's announced measurement results or not. If the error rate is higher than the threshold, the protocol will be aborted; otherwise, the protocol will be proceeded.

TP measures the particle pairs of $S_3^{''}$ corresponding to this Case with the $Z$ basis. Bob announces TP the corresponding $2L$ bits from $M_B$. TP calculates the error rate by judging whether her $Z$ basis measurement results on the corresponding particle pairs of $S_3^{''}$ are correctly correlated to the corresponding $2L$ bits from $M_B$ or not. If the error rate is higher than the threshold, the protocol will be aborted; otherwise, the protocol will be proceeded.

Alice publishes her measurement results on the $L$ particles of $S_3$ where she generated the fresh quantum states as found in Case ④. Bob publishes his measurement results on the corresponding $L$ particles of $S_3^{'}$. TP calculates the error rate by judging whether Alice's published measurement results are identical to Bob's published measurement results or not. If the error rate is higher than the threshold, the protocol will be aborted; otherwise, the protocol will be proceeded.

Step 6: The $L$ bits of $M_A$ used for generating the fresh quantum states by Alice in Case ④ are denoted as $K = \{k_1, k_2, \ldots, k_L\}$. Note that Bob can easily deduce $K$ from his measurement results on the corresponding particle of $S_3^{'}$. After that, Alice computes

$$g_i = p_a^i \oplus k_i \oplus m_a^i, \tag{9}$$

where the symbol $\oplus$ denotes the modulo 2 addition and $i = 1, 2, \ldots, L$. Similarly, Bob calculates

$$f_i = p_b^i \oplus k_i \oplus m_b^i, \tag{10}$$

where $i = 1, 2, \ldots, L$. Finally, Alice sends $g$ to TP, while Bob transmits $f$ to TP, where

$g=\{g_1,g_2,\ldots,g_L\}$ and $f=\{f_1,f_2,\ldots,f_L\}$.

Step 7: After receiving $g$ and $f$, TP computes

$$c_i=g_i\oplus f_i\oplus m_a^i\oplus m_b^i. \tag{11}$$

TP can determine whether $p_a^i$ and $p_b^i$ are equal or not according to $c_i$. When $c_i=0$, it can be derived that $p_a^i=p_b^i$; and when $c_i=1$, it can be derived that $p_a^i\neq p_b^i$. If $p_a^i=p_b^i$ for $i=1,2,\ldots,L$, TP will conclude that $p_a=p_b$; otherwise, TP will conclude that $p_a\neq p_b$. Finally, TP publishes the final comparison result between $p_a$ and $p_b$ to Alice and Bob, respectively.

Table 2 The usages of particles from $S_3$ under different Cases

| Case | The mode of Alice | The mode of Bob | Usage |
|---|---|---|---|
| Case ① | The REFLECT_A mode | The REFLECT_B mode | Eavesdropping detection |
| Case ② | The REFLECT_A mode | The MEASURE_B2 mode | Eavesdropping detection |
| Case ③ | The MEASURE_A2 mode | The REFLECT_B mode | Eavesdropping detection |
| Case ④ | The MEASURE_A2 mode | The MEASURE_B2 mode | Eavesdropping detection and private comparison |

## 3 Correctness analysis

In the following, we will validate that the output correctness of the proposed SQPC protocol can be guaranteed. After inserting Eq. (9) and Eq.(10) into Eq.(11), it can be easily obtained that

$$\begin{aligned} c_i &= g_i\oplus f_i\oplus m_a^i\oplus m_b^i \\ &=\left(p_a^i\oplus k_i\oplus m_a^i\right)\oplus\left(p_b^i\oplus k_i\oplus m_b^i\right)\oplus m_a^i\oplus m_b^i \\ &= p_a^i\oplus p_b^i. \end{aligned} \tag{12}$$

Based on Eq.(12), we can conclude that if $c_i=0$, it will have $p_a^i=p_b^i$; otherwise, it will have $p_a^i\neq p_b^i$. In other words, the output of the proposed SQPC protocol is correct.

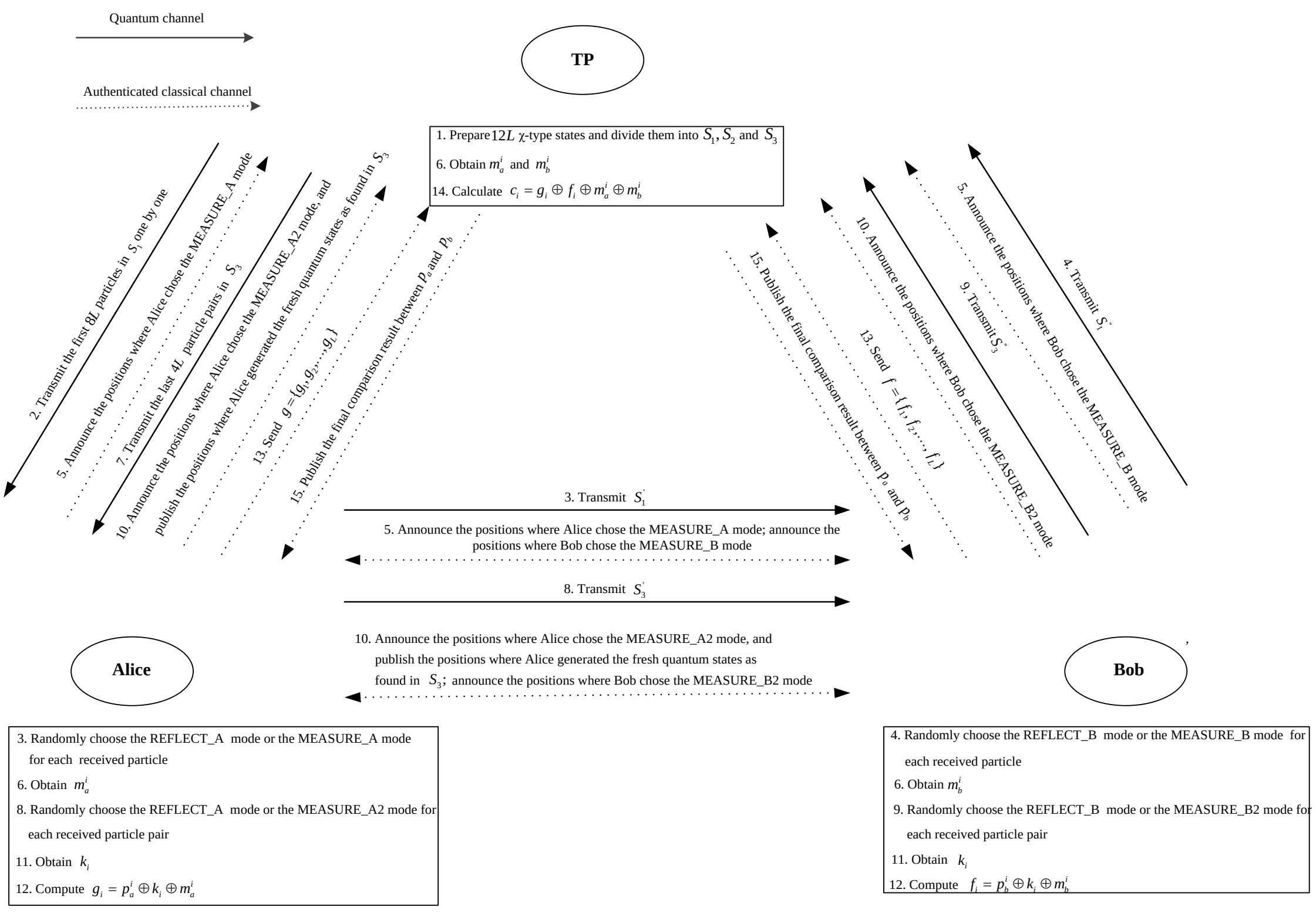

Fig.1 The flow chart of the proposed SQPC protocol without the security check processes

# 4 Security analysis

## 4.1 Outside attacks

In the proposed SQPC protocol, TP sends out the first $8L$ particles of $S_1$ and the last $4L$ particle pairs of $S_3$ in turn, so we should analyze the security against the transmissions of both the first $8L$ particles of $S_1$ and the last $4L$ particle pairs of $S_3$.

### 4.1.1 Outside attacks on the first $8L$ particles of $S_1$

(1) The intercept-resend attack

In the following, three different kinds of intercept-resend attack will be discussed in detail.

Firstly, Eve intercepts one of the first $8L$ particles in $S_1$ transmitted from TP and sends the fake one she has prepared within the $Z$ basis beforehand to Alice in Step 1; after Alice performs her operation on the fake particle, Eve intercepts the corresponding particle of $S_1^{'}$ and transmits the intercepted genuine one of $S_1$ to Bob in Step 2. When Alice chooses the REFLECT_A mode, no matter what mode Bob chooses, the presence of Eve cannot be discovered in Step 3; when Alice chooses the MEASURE_A mode and Bob chooses the MEASURE_B mode, the presence of Eve can be detected with the probability of $\frac{1}{2}$ in Case 2 of Step 3; when Alice chooses the MEASURE_A mode and Bob chooses the REFLECT_B mode, the probability that the intercepted particle is chosen for security check is $\frac{1}{2}$, and the probability that Alice's measurement result on this fake particle is not same to TP's measurement result on the corresponding particle of $S_1^{''}$ or not correctly correlated to TP's Bell basis measurement result on the corresponding particle pair of $S_3$ and TP's $Z$ basis measurement result on the corresponding particle of $S_2$ is also $\frac{1}{2}$, so the presence of Eve can be detected with the probability of $\frac{1}{2}\times\frac{1}{2}=\frac{1}{4}$ in Case 4 of Step 3. Hence, the probability that Eve's this intercept-resend attack on one of the first $8L$ particles in $S_1$ can be detected is $\frac{1}{4}\times 0+\frac{1}{4}\times 0+\frac{1}{4}\times\frac{1}{2}+\frac{1}{4}\times\frac{1}{4}=\frac{3}{16}$. It can be concluded that Eve's this intercept-resend attack on the first $8L$ particles in $S_1$ can be detected with the probability $1-\left(\frac{13}{16}\right)^{8L}$, which will be close to 1 if $L$ is large enough.

Secondly, Eve intercepts one of the first $8L$ particles in $S_1$ transmitted from TP and sends the fake one she has produced within the $Z$ basis in advance to Alice in Step 1; after Alice and Bob execute their operations, Eve intercepts the corresponding particle of $S_1^{''}$ and transmits the intercepted genuine one of $S_1$ to TP in Step 2. When Alice chooses the REFLECT_A mode and Bob chooses the REFLECT_B mode, the existence of Eve cannot be detected in Case 1 of Step 3; when Alice chooses the MEASURE_A mode and Bob chooses the MEASURE_B mode, the existence of Eve can be detected with the probability of $\frac{1}{2}$ in Case 2 of Step 3; when Alice chooses the REFLECT_A mode and Bob chooses the MEASURE_B mode, the probability that the intercepted particle is chosen for security check is $\frac{1}{2}$, and the probability that Bob's measurement result on the fake particle is not same to TP's measurement result on the corresponding particle of $S_1^{'}$ or not correctly correlated to TP's Bell basis measurement result on the corresponding particle pair of $S_3$ and TP's $Z$ basis measurement result on the corresponding particle of $S_2$ is also $\frac{1}{2}$, so the presence of Eve can be detected with the probability of $\frac{1}{2}\times\frac{1}{2}=\frac{1}{4}$ in Case 3 of Step 3; when Alice chooses the MEASURE_A mode and Bob chooses the REFLECT_B mode, the presence of Eve can be also discovered with the probability of $\frac{1}{2}\times\frac{1}{2}=\frac{1}{4}$ in Case 4 of Step 3. Consequently, Eve's this intercept-resend attack on one of the first $8L$ particles in $S_1$ can be detected with the probability $\frac{1}{4}\times 0+\frac{1}{4}\times\frac{1}{2}+\frac{1}{4}\times\frac{1}{4}+\frac{1}{4}\times\frac{1}{4}=\frac{1}{4}$. As a result, the probability that Eve's this intercept-resend attack on the first $8L$ particles in $S_1$ can be detected is $1-\left(\frac{3}{4}\right)^{8L}$, which will approach 1 if $L$ is large enough.

Thirdly, Eve intercepts the particle of $S_1^{'}$ transmitted from Alice and sends Bob the fake one she has already prepared within the $Z$ basis in Step 2; after Bob executes his operation on the fake particle, Eve intercepts the corresponding particle of $S_1^{''}$ and transmits the intercepted genuine one of $S_1^{'}$ to TP in Step 2. When Alice chooses the REFLECT_A mode and Bob chooses the REFLECT_B mode, the trace of Eve cannot be detected in Case 1 of Step 3; when Alice chooses the MEASURE_A mode and Bob chooses the MEASURE_B mode, the probability that Bob's measurement result on the fake particle is not same to Alice's measurement result on the genuine particle of $S_1$ and TP's measurement result on the corresponding particle of $S_1^{'}$ is $\frac{1}{2}$, so the presence of Eve can be detected with the probability of $\frac{1}{2}$ in Case 2 of Step 3; when Alice chooses the REFLECT_A mode and Bob chooses the MEASURE_B mode, the probability that the intercepted particle is chosen for security check is $\frac{1}{2}$, and the probability that Bob's measurement result on the fake particle is not same to TP's measurement result on the genuine particle of $S_1^{'}$ or not correctly correlated to TP's Bell basis measurement results on the corresponding particle pair of $S_3$ and TP's $Z$ basis measurement result on the corresponding particle of $S_2$ is $\frac{1}{2}$, so the presence of Eve can be detected with the probability of $\frac{1}{2}\times\frac{1}{2}=\frac{1}{4}$ in Case 3 of Step 3; when Alice chooses the MEASURE_A mode and Bob chooses the REFLECT_B mode, the probability that the eavesdropping behavior of Eve can be discovered is 0 in Case 4 of Step 3. Consequently, the probability that Eve's this intercept-resend attack on one of the first $8L$ particles in $S_1$ can be detected is $\frac{1}{4}\times 0+\frac{1}{4}\times\frac{1}{2}+\frac{1}{4}\times\frac{1}{4}+\frac{1}{4}\times 0=\frac{3}{16}$. It can be concluded that Eve's this intercept-resend attack on the first $8L$ particles in $S_1$ can be detected with the probability $1-\left(\frac{13}{16}\right)^{8L}$, which will converge to 1 if $L$ is large enough.

(2) The measure-resend attack

In order to obtain $m_a$ and $m_b$, Eve may intercept the first $8L$ particles in $S_1$ sent out from TP, employ the $Z$ basis to measure them and transmit the resulted states to Alice. When Alice enters into the MEASURE_A mode and Bob enters into the MEASURE_B mode, or when Alice enters into the MEASURE_A mode and Bob enters into the REFLECT_B mode, or when Alice enters into the REFLECT_A mode and Bob enters into the MEASURE_B mode, the presence of Eve cannot be detected in Step 3. Considering that Alice enters into the REFLECT_A mode and Bob enters into the REFLECT_B mode, Eve's $Z$ basis measurements damage the entanglement correlation of different particles within a $\chi$-type state, which makes her attack behavior be discovered undoubtedly.

In order to know $m_a$ and $m_b$, Eve may intercept the particles of $S_1^{'}$ sent out from Alice, perform the $Z$ basis measurements on them and send the resulted states to Bob. When Alice chooses the MEASURE_A mode and Bob chooses the MEASURE_B mode, or when Alice chooses the MEASURE_A mode and Bob chooses the REFLECT_B mode, or when Alice chooses the REFLECT_A mode and Bob chooses the MEASURE_B mode, Eve's attacks cannot be discovered in Step 3. But when Alice chooses the REFLECT_A mode and Bob chooses the REFLECT_B mode, Eve's $Z$ basis measurements destroy the entanglement correlation of different particles within a χ-type state; as a result, the presence of Eve can be detected inevitably.

In order to get $m_a$ and $m_b$ , Eve may intercept the particles of $S_1^*$ sent out from Bob, utilize the $Z$ basis to measure them and transmit the resulted states to TP . When Alice comes into the MEASURE_A mode and Bob comes into the MEASURE_B mode, or when Alice comes into the MEASURE_A mode and Bob comes into the REFLECT_B mode, or when Alice comes into the REFLECT_A mode and Bob comes into the MEASURE_B mode, Eve's existence cannot be perceived in Step 3. However, when Alice comes into the REFLECT_A mode and Bob comes into the REFLECT_B mode, Eve's $Z$ basis measurements destroy the entanglement correlation of different particles within a χ-type state; in this way, Eve's attack can be caught definitely.

(3) The entangle-measure attack

Eve may launch her entangle-measure attack on the transmitted particles by adopting two unitaries, $U_E$ and $U_F$ , where $U_E$ and $U_F$ share a common probe space with the auxiliary state $|E\rangle$ . There are three different types of entangle-measure attack: ①Eve may perform $U_E$ on the qubits transmitted from TP to Alice in Step 1 and perform $U_F$ on the qubits transmitted from Alice to Bob in Step 2; ② Eve may implement $U_E$ on the qubits transmitted from TP to Alice in Step 1 and implement $U_F$ on the qubits transmitted from Bob to TP in Step 2; ③ Eve may implement $U_E$ on the qubits transmitted from Alice to Bob in Step 2 and perform $U_F$ on the qubits transmitted from Bob to TP in Step 2. In the following, we discuss the first kind of entangle-measure attack concretely, which can be depicted as Fig 2.

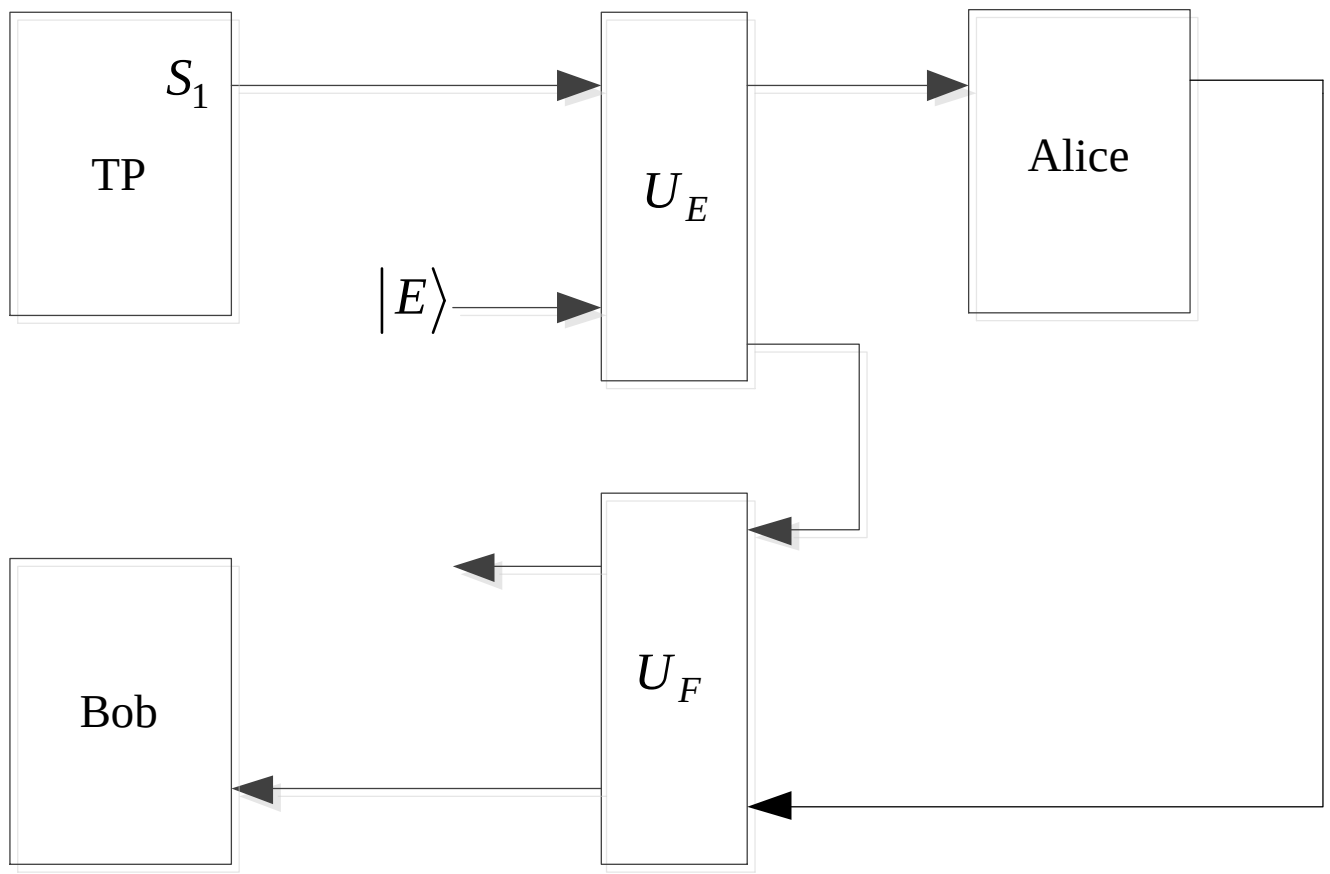

Fig.2 Eve's first kind of entangle-measure attack with $U_E$ and $U_F$

**Theorem 1.** *Assume that Eve implements $U_E$ on the qubit from TP to Alice in Step 1 and applies $U_F$ on the qubit from Alice to Bob in Step 2. For introducing no error in Step 3, the final state of Eve's probe should be independent of not only the operations of Alice and Bob but also their measurement results. As a result, Eve acquires no information about $m_a$ or $m_b$ .*

**Proof.** When $U_E$ is applied on a qubit prepared in the $Z$ basis, we have

$$U_E\left(|0\rangle|E\rangle\right)=\lambda_{00}|0\rangle|\varepsilon_{00}\rangle+\lambda_{01}|1\rangle|\varepsilon_{01}\rangle \tag{13}$$

and

$$U_E\left(|1\rangle|E\rangle\right)=\lambda_{10}|0\rangle|\varepsilon_{10}\rangle+\lambda_{11}|1\rangle|\varepsilon_{11}\rangle . \tag{14}$$

Here, $|\varepsilon_{00}\rangle$ , $|\varepsilon_{01}\rangle$ , $|\varepsilon_{10}\rangle$ and $|\varepsilon_{11}\rangle$ are Eve's probe states decided by $U_E$ , $|\lambda_{00}|^2+|\lambda_{01}|^2=1$ and $|\lambda_{10}|^2+|\lambda_{11}|^2=1$.

When Eve performs $U_E$ on the qubit of the first $8L$ particles in $S_1$ transmitted from TP to Alice in Step 1, according to Eq.(5), Eq.(13) and Eq.(14), we can get

$$\begin{aligned}
U_E\left(|\chi^{00}\rangle_{1234}|E\rangle\right) &= U_E\left[\frac{\sqrt{2}}{4}\left(|0000\rangle-|0101\rangle+|0011\rangle+|0110\rangle+|1001\rangle+|1010\rangle+|1100\rangle-|1111\rangle\right)_{1234}|E\rangle\right]\\
&=\frac{\sqrt{2}}{4}\Big[\left(\lambda_{00}|0\rangle_1|\varepsilon_{00}\rangle+\lambda_{01}|1\rangle_1|\varepsilon_{01}\rangle\right)|000\rangle_{234}-\left(\lambda_{00}|0\rangle_1|\varepsilon_{00}\rangle+\lambda_{01}|1\rangle_1|\varepsilon_{01}\rangle\right)|101\rangle_{234}\\
&\quad+\left(\lambda_{00}|0\rangle_1|\varepsilon_{00}\rangle+\lambda_{01}|1\rangle_1|\varepsilon_{01}\rangle\right)|011\rangle_{234}+\left(\lambda_{00}|0\rangle_1|\varepsilon_{00}\rangle+\lambda_{01}|1\rangle_1|\varepsilon_{01}\rangle\right)|110\rangle_{234}\\
&\quad+\left(\lambda_{10}|0\rangle_1|\varepsilon_{10}\rangle+\lambda_{11}|1\rangle_1|\varepsilon_{11}\rangle\right)|001\rangle_{234}+\left(\lambda_{10}|0\rangle_1|\varepsilon_{10}\rangle+\lambda_{11}|1\rangle_1|\varepsilon_{11}\rangle\right)|010\rangle_{234}\\
&\quad+\left(\lambda_{10}|0\rangle_1|\varepsilon_{10}\rangle+\lambda_{11}|1\rangle_1|\varepsilon_{11}\rangle\right)|100\rangle_{234}-\left(\lambda_{10}|0\rangle_1|\varepsilon_{10}\rangle+\lambda_{11}|1\rangle_1|\varepsilon_{11}\rangle\right)|111\rangle_{234}\Big]\\
&=\frac{\sqrt{2}}{4}\Big(\lambda_{00}|0\rangle_1|000\rangle_{234}|\varepsilon_{00}\rangle+\lambda_{01}|1\rangle_1|000\rangle_{234}|\varepsilon_{01}\rangle-\lambda_{00}|0\rangle_1|101\rangle_{234}|\varepsilon_{00}\rangle-\lambda_{01}|1\rangle_1|101\rangle_{234}|\varepsilon_{01}\rangle\\
&\quad+\lambda_{00}|0\rangle_1|011\rangle_{234}|\varepsilon_{00}\rangle+\lambda_{01}|1\rangle_1|011\rangle_{234}|\varepsilon_{01}\rangle+\lambda_{00}|0\rangle_1|110\rangle_{234}|\varepsilon_{00}\rangle+\lambda_{01}|1\rangle_1|110\rangle_{234}|\varepsilon_{01}\rangle\\
&\quad+\lambda_{10}|0\rangle_1|001\rangle_{234}|\varepsilon_{10}\rangle+\lambda_{11}|1\rangle_1|001\rangle_{234}|\varepsilon_{11}\rangle+\lambda_{10}|0\rangle_1|010\rangle_{234}|\varepsilon_{10}\rangle+\lambda_{11}|1\rangle_1|010\rangle_{234}|\varepsilon_{11}\rangle\\
&\quad+\lambda_{10}|0\rangle_1|100\rangle_{234}|\varepsilon_{10}\rangle+\lambda_{11}|1\rangle_1|100\rangle_{234}|\varepsilon_{11}\rangle-\lambda_{10}|0\rangle_1|111\rangle_{234}|\varepsilon_{10}\rangle-\lambda_{11}|1\rangle_1|111\rangle_{234}|\varepsilon_{11}\rangle\Big)\\
&=\frac{\sqrt{2}}{4}\Big(|0\rangle_1|000\rangle_{234}|\partial_{00}\rangle+|1\rangle_1|000\rangle_{234}|\partial_{01}\rangle-|0\rangle_1|101\rangle_{234}|\partial_{00}\rangle-|1\rangle_1|101\rangle_{234}|\partial_{01}\rangle\\
&\quad+|0\rangle_1|011\rangle_{234}|\partial_{00}\rangle+|1\rangle_1|011\rangle_{234}|\partial_{01}\rangle+|0\rangle_1|110\rangle_{234}|\partial_{00}\rangle+|1\rangle_1|110\rangle_{234}|\partial_{01}\rangle\\
&\quad+|0\rangle_1|001\rangle_{234}|\partial_{10}\rangle+|1\rangle_1|001\rangle_{234}|\partial_{11}\rangle+|0\rangle_1|010\rangle_{234}|\partial_{10}\rangle+|1\rangle_1|010\rangle_{234}|\partial_{11}\rangle\\
&\quad+|0\rangle_1|100\rangle_{234}|\partial_{10}\rangle+|1\rangle_1|100\rangle_{234}|\partial_{11}\rangle-|0\rangle_1|111\rangle_{234}|\partial_{10}\rangle-|1\rangle_1|111\rangle_{234}|\partial_{11}\rangle\Big),
\end{aligned} \tag{15}$$

where $\partial_{00}=\lambda_{00}|\varepsilon_{00}\rangle$ , $\partial_{01}=\lambda_{01}|\varepsilon_{01}\rangle$ , $\partial_{10}=\lambda_{10}|\varepsilon_{10}\rangle$ and $\partial_{11}=\lambda_{11}|\varepsilon_{11}\rangle$ .

(i) Consider the case that Alice chooses the MEASURE_A mode and Bob chooses the

MEASURE_B mode. In order to escape the security checks in Case 2 of Step 3, Eve should make TP's Bell basis measurement result on the corresponding particle pair of $S_3$ and $Z$ basis measurement result on the corresponding particle of $S_2$ correctly correlate to Alice's measurement result on the particle of $S_1$, in accordance with Eq.(6). Hence, it should satisfy that

$$\left|\partial_{01}\right\rangle = \left|\partial_{10}\right\rangle = 0 . \tag{16}$$

Inserting Eq.(16) into Eq.(15) generates

$$\begin{aligned} U_E\left(\left|\chi^{00}\right\rangle_{1234}\left|E\right\rangle\right) &= \frac{\sqrt{2}}{4}\Big(\left|0\right\rangle_1\left|000\right\rangle_{234}\left|\partial_{00}\right\rangle - \left|0\right\rangle_1\left|101\right\rangle_{234}\left|\partial_{00}\right\rangle + \left|0\right\rangle_1\left|011\right\rangle_{234}\left|\partial_{00}\right\rangle + \left|0\right\rangle_1\left|110\right\rangle_{234}\left|\partial_{00}\right\rangle \\ &\quad + \left|1\right\rangle_1\left|001\right\rangle_{234}\left|\partial_{11}\right\rangle + \left|1\right\rangle_1\left|010\right\rangle_{234}\left|\partial_{11}\right\rangle + \left|1\right\rangle_1\left|100\right\rangle_{234}\left|\partial_{11}\right\rangle - \left|1\right\rangle_1\left|111\right\rangle_{234}\left|\partial_{11}\right\rangle\Big) \\ &= \frac{1}{2}\Big(\left|0\right\rangle_1\left|0\right\rangle_2\left|\phi^+\right\rangle_{34}\left|\partial_{00}\right\rangle + \left|1\right\rangle_1\left|1\right\rangle_2\left|\phi^-\right\rangle_{34}\left|\partial_{11}\right\rangle - \left|0\right\rangle_1\left|1\right\rangle_2\left|\psi^-\right\rangle_{34}\left|\partial_{00}\right\rangle + \left|1\right\rangle_1\left|0\right\rangle_2\left|\psi^+\right\rangle_{34}\left|\partial_{11}\right\rangle\Big). \end{aligned} \tag{17}$$

After Alice performs the MEASURE_A mode, according to Eq.(17), the global state of the composite system is collapsed into $\left|0\right\rangle_1\left|0\right\rangle_2\left|\phi^+\right\rangle_{34}\left|\partial_{00}\right\rangle$, $\left|1\right\rangle_1\left|1\right\rangle_2\left|\phi^-\right\rangle_{34}\left|\partial_{11}\right\rangle$, $\left|0\right\rangle_1\left|1\right\rangle_2\left|\psi^-\right\rangle_{34}\left|\partial_{00}\right\rangle$ or $\left|1\right\rangle_1\left|0\right\rangle_2\left|\psi^+\right\rangle_{34}\left|\partial_{11}\right\rangle$. When Bob performs the MEASURE_B mode, in order to make Alice's measurement result on the particle of $S_1$ same to Bob's measurement result on the corresponding particle of $S_1^{'}$, the whole quantum system after being applied with $U_F$ should be

$$U_F\left(\left|0\right\rangle_1\left|0\right\rangle_2\left|\phi^+\right\rangle_{34}\left|\partial_{00}\right\rangle\right) = \left|0\right\rangle_1\left|0\right\rangle_2\left|\phi^+\right\rangle_{34}\left|F_{00}\right\rangle, \tag{18}$$

$$U_F\left(\left|1\right\rangle_1\left|1\right\rangle_2\left|\phi^-\right\rangle_{34}\left|\partial_{11}\right\rangle\right) = \left|1\right\rangle_1\left|1\right\rangle_2\left|\phi^-\right\rangle_{34}\left|F_{11}\right\rangle, \tag{19}$$

$$U_F\left(\left|0\right\rangle_1\left|1\right\rangle_2\left|\psi^-\right\rangle_{34}\left|\partial_{00}\right\rangle\right) = \left|0\right\rangle_1\left|1\right\rangle_2\left|\psi^-\right\rangle_{34}\left|F_{00}\right\rangle \tag{20}$$

or

$$U_F\left(\left|1\right\rangle_1\left|0\right\rangle_2\left|\psi^+\right\rangle_{34}\left|\partial_{11}\right\rangle\right) = \left|1\right\rangle_1\left|0\right\rangle_2\left|\psi^+\right\rangle_{34}\left|F_{11}\right\rangle . \tag{21}$$

which means that $U_F$ cannot alter the state of the corresponding particle of $S_1^{'}$ from Alice. Otherwise, Eve can be discovered with a non-zero probability.

(ii) Consider the case that Alice chooses the MEASURE_A mode and Bob chooses the REFLECT_B mode. After Alice performs the MEASURE_A mode, according to Eq.(17), the global state of the composite system is collapsed into $\left|0\right\rangle_1\left|0\right\rangle_2\left|\phi^+\right\rangle_{34}\left|\partial_{00}\right\rangle$, $\left|1\right\rangle_1\left|1\right\rangle_2\left|\phi^-\right\rangle_{34}\left|\partial_{11}\right\rangle$, $\left|0\right\rangle_1\left|1\right\rangle_2\left|\psi^-\right\rangle_{34}\left|\partial_{00}\right\rangle$ or $\left|1\right\rangle_1\left|0\right\rangle_2\left|\psi^+\right\rangle_{34}\left|\partial_{11}\right\rangle$. Hence, after Eve imposes $U_F$ and Bob performs the REFLECT_B mode, by virtue of Eqs.(18-21), TP's Bell basis measurement result on the corresponding particle pair of $S_3$ and $Z$ basis measurement result on the corresponding particle of $S_2$ are automatically correctly correlated to Alice's measurement result on the particle of $S_1$; and TP's $Z$ basis measurement on the corresponding particle of $S_1^{''}$ is naturally identical to Alice's measurement result on the particle of $S_1$. It can be concluded that as long as Eq.(16) and Eqs.(18-21) are satisfied, Eve cannot be discovered in Case 4 of Step 3.

(iii) Consider the case that Alice chooses the REFLECT_A mode and Bob chooses the MEASURE_B mode. After Alice performs the REFLECT_A mode, on the basis of Eq.(17) and Eqs.(18-21), the whole quantum system after being applied with $U_F$ should be

$$U_F\left[U_E\left(\left|\chi^{00}\right\rangle_{1234}|E\rangle\right)\right]$$
$$=U_F\left[\frac{1}{2}\left(|0\rangle_1|0\rangle_2\left|\phi^+\right\rangle_{34}|\partial_{00}\rangle+|1\rangle_1|1\rangle_2\left|\phi^-\right\rangle_{34}|\partial_{11}\rangle-|0\rangle_1|1\rangle_2\left|\psi^-\right\rangle_{34}|\partial_{00}\rangle+|1\rangle_1|0\rangle_2\left|\psi^+\right\rangle_{34}|\partial_{11}\rangle\right)\right]$$
$$=\frac{1}{2}\left(|0\rangle_1|0\rangle_2\left|\phi^+\right\rangle_{34}|F_{00}\rangle+|1\rangle_1|1\rangle_2\left|\phi^-\right\rangle_{34}|F_{11}\rangle-|0\rangle_1|1\rangle_2\left|\psi^-\right\rangle_{34}|F_{00}\rangle+|1\rangle_1|0\rangle_2\left|\psi^+\right\rangle_{34}|F_{11}\rangle\right). \quad (22)$$

After Bob performs the MEASURE_B mode, according to Eq.(22), the global state of the composite system is collapsed into $|0\rangle_1|0\rangle_2\left|\phi^+\right\rangle_{34}|F_{00}\rangle$, $|1\rangle_1|1\rangle_2\left|\phi^-\right\rangle_{34}|F_{11}\rangle$, $|0\rangle_1|1\rangle_2\left|\psi^-\right\rangle_{34}|F_{00}\rangle$ or $|1\rangle_1|0\rangle_2\left|\psi^+\right\rangle_{34}|F_{11}\rangle$. Hence, TP's $Z$ basis measurement on the corresponding particle of $S_1^{''}$, TP's Bell basis measurement result on the corresponding particle pair of $S_3$ and TP's $Z$ basis measurement results on the corresponding particle of $S_2$ are correctly correlated to Bob's measurement result on the particle of $S_1^{'}$. As a result, the presence of Eve cannot be discovered in Case 3 of Step 3.

(iv) Consider the case that Alice chooses the REFLECT_A mode and Bob chooses the REFLECT_B mode. In order for Eve not being detected in Case 1 of Step 3, TP's $FMB$ basis measurement result should be $\left|\chi^{00}\right\rangle$. Hence, according to Eq.(22), it should satisfy

$$U_F\left[U_E\left(\left|\chi^{00}\right\rangle_{1234}|E\rangle\right)\right]$$
$$=\frac{1}{2}\left(|0\rangle_1|0\rangle_2\left|\phi^+\right\rangle_{34}|F_{00}\rangle+|1\rangle_1|1\rangle_2\left|\phi^-\right\rangle_{34}|F_{11}\rangle-|0\rangle_1|1\rangle_2\left|\psi^-\right\rangle_{34}|F_{00}\rangle+|1\rangle_1|0\rangle_2\left|\psi^+\right\rangle_{34}|F_{11}\rangle\right)$$
$$=\frac{1}{2}\left(|0\rangle_1|0\rangle_2\left|\phi^+\right\rangle_{34}+|1\rangle_1|1\rangle_2\left|\phi^-\right\rangle_{34}-|0\rangle_1|1\rangle_2\left|\psi^-\right\rangle_{34}+|1\rangle_1|0\rangle_2\left|\psi^+\right\rangle_{34}\right)|F\rangle$$
$$=\left|\chi^{00}\right\rangle_{1234}|F\rangle. \quad (23)$$

We can deduce from Eq.(23) that

$$|F_{00}\rangle=|F_{11}\rangle=|F\rangle. \quad (24)$$

(v) Applying Eq.(24) into Eqs.(18-21) produces

$$U_F\left(|0\rangle_1|0\rangle_2\left|\phi^+\right\rangle_{34}|\partial_{00}\rangle\right)=|0\rangle_1|0\rangle_2\left|\phi^+\right\rangle_{34}|F\rangle, \quad (25)$$

$$U_F\left(|1\rangle_1|1\rangle_2|\phi^-\rangle_{34}|\partial_{11}\rangle\right)=|1\rangle_1|1\rangle_2|\phi^-\rangle_{34}|F\rangle, \quad (26)$$

$$U_F\left(|0\rangle_1|1\rangle_2|\psi^-\rangle_{34}|\partial_{00}\rangle\right)=|0\rangle_1|1\rangle_2|\psi^-\rangle_{34}|F\rangle \quad (27)$$

and

$$U_F\left(|1\rangle_1|0\rangle_2|\psi^+\rangle_{34}|\partial_{11}\rangle\right)=|1\rangle_1|0\rangle_2|\psi^+\rangle_{34}|F\rangle. \quad (28)$$

According to Eqs.(23,25-28), we can conclude that when Eve implements $U_E$ on the qubit from TP to Alice in Step 1 and applies $U_F$ on the qubit from Alice to Bob in Step 2, for introducing no error in Step 3, the final state of Eve's probe should be independent of not only the operations of Alice and Bob but also their measurement results. As a result, Eve acquires no information about $m_a$ or $m_b$.

In addition, it can be also verified by using the similar method that Eve cannot acquire any useful information either in the other foregoing two kinds of entangle-measure attack.

(4) The Trojan horse attacks

Because the particles of $S_1$ travel from TP to Alice, from Alice to Bob and back from Bob to TP, the Trojan horse attacks from Eve should be considered. To overcome the invisible photon eavesdropping attack, the receiver can put a wavelength filter in front of her device to eliminate the illegitimate photon signal [34,35]. To resist the delay-photon Trojan horse attack, the receiver can employ a photon number splitter (PNS: 50/50) to split each sample signal into two parts and calculate the multiphoton rate after using the correct measuring bases to measure the resulted signals [34,35].

**4.1.2 Outside attacks on the last $4L$ particle pairs of $S_3$**

(1) The intercept-resend attack

We analyze three different kinds of intercept-resend attack as follows.

Firstly, in Step 4, Eve intercepts one of the last $4L$ particle pairs of $S_3$ transmitted from TP and sends the fake particle pair she has already prepared in the $Z$ basis to Alice; after Alice executes her operation on the received fake particle pair, Eve intercepts the corresponding particle pair of $S_3^{'}$ and sends the intercepted genuine particle pair of $S_3$ to Bob. When Alice chooses the REFLECT_A mode, no matter what mode Bob chooses, the eavesdropping behavior of Eve cannot be discovered in Step 5; when Alice chooses the MEASURE_A2 mode and Bob chooses the REFLECT_B mode, the probability that Alice's measurement result on the fake particle pair is not correctly correlated to TP's Bell basis measurement result on the corresponding particles of $S_1$ and $S_2$ is $\frac{3}{4}$, so the presence of Eve can be detected with the probability $\frac{3}{4}$ in Case③of Step 5; when Alice chooses the MEASURE_A2 mode and Bob chooses the MEASURE_B2 mode, the probability that Alice's measurement result on the fake particle pair is not correctly correlated to TP's Bell basis measurement result on the corresponding particles of $S_1$ and $S_2$ is $\frac{3}{4}$, so the existence of Eve can be discovered with the probability $\frac{3}{4}$ in Case ④ of Step 5. Hence, Eve's this intercept-resend attack on one of the last $4L$ particle pairs of $S_3$ can be detected with the probability $\frac{1}{4}\times 0+\frac{1}{4}\times 0+\frac{1}{4}\times\frac{3}{4}+\frac{1}{4}\times\frac{3}{4}=\frac{3}{8}$. It can be concluded that Eve's this intercept-resend attack on the last $4L$ particle pairs of $S_3$ is $1-\left(\frac{5}{8}\right)^{4L}$, which will converge to 1 if $L$ is large enough.

Secondly, in Step 4, Eve intercepts one of the last $4L$ particle pairs of $S_3$ transmitted from TP and sends Alice the fake particle pair she has prepared beforehand in the $Z$ basis; after Alice and Bob perform their operation mode, Eve intercepts the particle pair of $S_3^{''}$ and transmits the intercepted genuine particle pair of $S_3$ to TP. When Alice chooses the REFLECT_A mode and Bob chooses the REFLECT_B mode, Eve's attack behavior cannot be discovered in Case ① of Step 5; when Alice chooses the REFLECT_A mode and Bob chooses the MEASURE_B2 mode, the probability that Bob's measurement result on the fake particle pair is not correctly correlated to TP's Bell basis measurement result on the corresponding particles of $S_1$ and $S_2$ is $\frac{3}{4}$, and the probability that TP's measurement result on the intercepted genuine particle pair of $S_3$ is not same to the corresponding bit values of $M_B$ is also $\frac{3}{4}$, so the attack behavior of Eve can be detected with the probability of $\frac{3}{4}$ in Case ② of Step 5; when Alice chooses the MEASURE_A2 mode and Bob chooses the REFLECT_B mode, the probability that Alice's measurement result on the fake particle pair is not correctly correlated to TP's Bell basis measurement result on the corresponding particles of $S_1$ and $S_2$ is $\frac{3}{4}$, so the eavesdropping behavior of Eve can be discovered with the probability of $\frac{3}{4}$ in Case ③ of Step 5; when Alice chooses the MEASURE_A2 mode and Bob chooses the MEASURE_B2 mode, the probability that Alice's measurement result on the fake particle pair is not correctly correlated to TP's Bell basis measurement result on the corresponding particles of $S_1$ and $S_2$ is $\frac{3}{4}$, and the probability that TP's measurement result on the intercepted genuine particle pair of $S_3$ is not same to the corresponding bit values of $M_B$ is $\frac{3}{4}$, so the probability that the presence of Eve can be discovered is also $\frac{3}{4}$ in Case ④ of Step 5. Consequently, the probability that Eve's this intercept-resend attack on one of the last $4L$ particle pairs of $S_3$ can be detected is $\frac{1}{4}\times 0+\frac{1}{4}\times\frac{3}{4}+\frac{1}{4}\times\frac{3}{4}+\frac{1}{4}\times\frac{3}{4}=\frac{9}{16}$. As a result, the probability that Eve's this intercept-resend attack on the last $4L$ particle pairs of $S_3$ can be detected is $1-\left(\frac{7}{16}\right)^{4L}$, which will approach 1 if $L$ is large enough.

Thirdly, in Step 4, Eve intercepts the particle pair of $S_3^{'}$ transmitted from Alice and sends Bob the fake particle pair she has already prepared in the $Z$ basis; after Bob performs his operation on the received fake particle pair, Eve intercepts the particle pair of $S_3^{''}$ and transmits the intercepted genuine particle pair of $S_3^{'}$ to TP. When Alice chooses the REFLECT_A mode and Bob chooses the REFLECT_B mode, Eve's eavesdropping behavior cannot be discovered in Case ① of Step 5; when Alice chooses the REFLECT_A mode and Bob chooses the MEASURE_B2 mode, the probability that Bob's measurement result on the fake particle pair is not correctly correlated to TP's Bell basis measurement result on the corresponding particles of $S_1$ and $S_2$ is $\frac{3}{4}$, and the probability that TP's measurement result on the intercepted genuine particle pair of $S_3^{'}$ is not same to the corresponding bit values of $M_B$ is $\frac{3}{4}$, so the probability that the presence of Eve can be discovered is $\frac{3}{4}$ in Case ② of Step 5; when Alice chooses the MEASURE_A2 mode and Bob chooses the REFLECT_B mode, the attack behavior of Eve cannot be detected in Case ③ of Step 5; when Alice chooses the MEASURE_A2 mode and Bob chooses the MEASURE_B2 mode, the probability that TP's measurement result on the intercepted genuine particle pair of $S_3^{'}$ is not same to the corresponding bit values of $M_B$ is $\frac{3}{4}$, and the probability that Bob's measurement result on the fake particle pair is not same to the state of fresh particle pair Alice generates as found is also $\frac{3}{4}$, so the presence of Eve can be discovered with the probability of $\frac{3}{4}$ in Case ④ of Step 5. Hence, the probability that Eve's this intercept-resend attack on one of the last $4L$ particle pairs of $S_3$ can be detected is $\frac{1}{4}\times 0+\frac{1}{4}\times\frac{3}{4}+\frac{1}{4}\times 0+\frac{1}{4}\times\frac{3}{4}=\frac{3}{8}$. As a result, we can conclude that Eve's this intercept-resend attack on the last $4L$ particle pairs of $S_3$ can be detected with the probability $1-\left(\frac{5}{8}\right)^{4L}$, which will converge to 1 if $L$ is large enough.

(2) The measure-resend attack

In order to get $K$, Eve may intercept the last $4L$ particle pairs of $S_3$ sent out from TP, employ the $Z$ basis to measure them and transmit the resulted states to Alice. When Alice enters into the MEASURE_A2 mode and Bob enters into the MEASURE_B2 mode, or when Alice enters into the MEASURE_A2 mode and Bob enters into the REFLECT_B mode, or when Alice enters into the REFLECT_A mode and Bob enters into the MEASURE_B2 mode, the presence of Eve cannot be detected in Step 5. Considering that Alice executes the REFLECT_A mode and Bob executes the REFLECT_B mode, Eve's $Z$ basis measurements damage the entanglement correlation of different particles within a $\chi$-type state, which makes her attack behavior be discovered undoubtedly.

In order to get $K$ , Eve may intercept the particle pairs of $S_3^{'}$ sent out from Alice, employ the $Z$ basis to measure them and transmit the resulted states to Bob. When Alice enters into the MEASURE_A2 mode and Bob enters into the MEASURE_B2 mode, or when Alice enters into the MEASURE_A2 mode and Bob enters into the REFLECT_B mode, or when Alice enters into the REFLECT_A mode and Bob enters into the MEASURE_B2 mode, the presence of Eve cannot be detected in Step 5. Considering that Alice enters into the REFLECT_A mode and Bob enters into the REFLECT_B mode, Eve's $Z$ basis measurements destroy the entanglement correlation of different particles within a $\chi$-type state; as a result, the presence of Eve can be detected inevitably.

In order to get $K$ , Eve may intercept the particle pairs of $S_3^{''}$ sent out from Bob, employ the $Z$ basis to measure them and transmit the resulted states to TP. When Alice enters into the MEASURE_A2 mode and Bob enters into the MEASURE_B2 mode, or when Alice enters into the MEASURE_A2 mode and Bob enters into the REFLECT_B mode, or when Alice enters into the REFLECT_A mode and Bob enters into the MEASURE_B2 mode, the presence of Eve cannot be detected in Step 5. Considering that Alice comes into the REFLECT_A and Bob comes into the REFLECT_B mode, Eve's $Z$ basis measurements destroy the entanglement correlation of different particles within a $\chi$-type state; in this way, Eve's attack can be caught definitely.

(3) The entangle-measure attack

Eve may impose her entangle-measure attack on the transmitted particle pair with two unitaries, $U_E$ and $U_F$ , where $U_E$ and $U_F$ share a common probe space with the auxiliary state $|\Omega\rangle$ . In order to obtain $K$ , if Eve is clever enough, she will launch her entangle-measure attack as follows, which can be depicted as Fig.3: she performs $U_E$ on the particle pair transmitted from TP to Alice and $U_F$ on the particle pair transmitted from Alice to Bob in Step 4.

**Theorem 2.** *Assume that Eve performs* $U_E$ *on the particle pair from TP to Alice and* $U_F$ *on the particle pair from Alice to Bob in Step 4. For introducing no error in Step 5, the final state of Eve's probe should be independent of not only the operations of Alice and Bob but also their measurement results. As a result, Eve acquires no information about* $K$ .

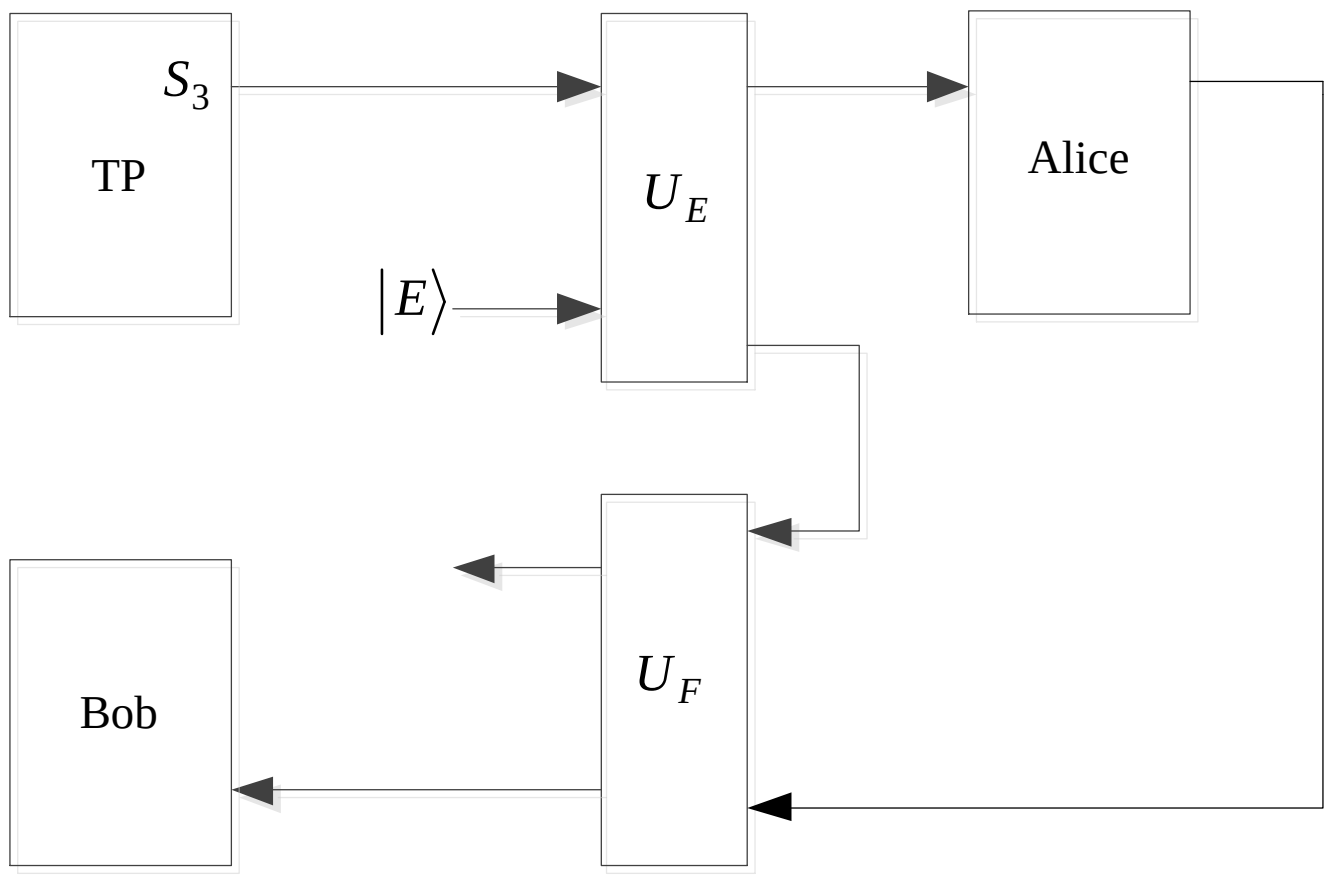

Fig.3 Eve's entangle-measure attack with $U_E$ and $U_F$

**Proof.** When a qubit within the $Z$ basis is performed with $U_E$, we have

$$U_E\left(|0\rangle|\Omega\rangle\right)=\delta_{00}|0\rangle|\varpi_{00}\rangle+\delta_{01}|1\rangle|\varpi_{01}\rangle \tag{29}$$

and

$$U_E\left(|1\rangle|\Omega\rangle\right)=\delta_{10}|0\rangle|\varpi_{10}\rangle+\delta_{11}|1\rangle|\varpi_{11}\rangle. \tag{30}$$

Here, $|\varpi_{00}\rangle$, $|\varpi_{01}\rangle$, $|\varpi_{10}\rangle$ and $|\varpi_{11}\rangle$ are Eve's probe states determined by $U_E$, $|\delta_{00}|^2+|\delta_{01}|^2=1$ and $|\delta_{10}|^2+|\delta_{11}|^2=1$.

When Eve performs $U_E$ on the particle pair of $S_3$ transmitted from TP to Alice in Step 4, we can get

$$U_E\left(|\chi^{00}\rangle_{1234}|\Omega\rangle\right)=U_E\left[\frac{\sqrt{2}}{4}\left(|0000\rangle-|0101\rangle+|0011\rangle+|0110\rangle+|1001\rangle+|1010\rangle+|1100\rangle-|1111\rangle\right)_{1234}|\Omega\rangle\right]$$

$$\begin{aligned}
=\frac{\sqrt{2}}{4}\Big[&|00\rangle_{12}\left(\delta_{00}|0\rangle_3|\varpi_{00}\rangle+\delta_{01}|1\rangle_3|\varpi_{01}\rangle\right)\left(\delta_{00}|0\rangle_4|\varpi_{00}\rangle+\delta_{01}|1\rangle_4|\varpi_{01}\rangle\right)\\
&-|01\rangle_{12}\left(\delta_{00}|0\rangle_3|\varpi_{00}\rangle+\delta_{01}|1\rangle_3|\varpi_{01}\rangle\right)\left(\delta_{10}|0\rangle_4|\varpi_{10}\rangle+\delta_{11}|1\rangle_4|\varpi_{11}\rangle\right)\\
&+|00\rangle_{12}\left(\delta_{10}|0\rangle_3|\varpi_{10}\rangle+\delta_{11}|1\rangle_3|\varpi_{11}\rangle\right)\left(\delta_{10}|0\rangle_4|\varpi_{10}\rangle+\delta_{11}|1\rangle_4|\varpi_{11}\rangle\right)\\
&+|01\rangle_{12}\left(\delta_{10}|0\rangle_3|\varpi_{10}\rangle+\delta_{11}|1\rangle_3|\varpi_{11}\rangle\right)\left(\delta_{00}|0\rangle_4|\varpi_{00}\rangle+\delta_{01}|1\rangle_4|\varpi_{01}\rangle\right)\\
&+|10\rangle_{12}\left(\delta_{00}|0\rangle_3|\varpi_{00}\rangle+\delta_{01}|1\rangle_3|\varpi_{01}\rangle\right)\left(\delta_{10}|0\rangle_4|\varpi_{10}\rangle+\delta_{11}|1\rangle_4|\varpi_{11}\rangle\right)\\
&+|10\rangle_{12}\left(\delta_{10}|0\rangle_3|\varpi_{10}\rangle+\delta_{11}|1\rangle_3|\varpi_{11}\rangle\right)\left(\delta_{00}|0\rangle_4|\varpi_{00}\rangle+\delta_{01}|1\rangle_4|\varpi_{01}\rangle\right)\\
&+|11\rangle_{12}\left(\delta_{00}|0\rangle_3|\varpi_{00}\rangle+\delta_{01}|1\rangle_3|\varpi_{01}\rangle\right)\left(\delta_{00}|0\rangle_4|\varpi_{00}\rangle+\delta_{01}|1\rangle_4|\varpi_{01}\rangle\right)\\
&-|11\rangle_{12}\left(\delta_{10}|0\rangle_3|\varpi_{10}\rangle+\delta_{11}|1\rangle_3|\varpi_{11}\rangle\right)\left(\delta_{10}|0\rangle_4|\varpi_{10}\rangle+\delta_{11}|1\rangle_4|\varpi_{11}\rangle\right)\Big]
\end{aligned}$$

$$
\begin{aligned}
=\frac{\sqrt{2}}{4}\Big[&|00\rangle_{12}|0\rangle_3|0\rangle_4\left(\delta_{00}^2|\varpi_{00}\rangle|\varpi_{00}\rangle+\delta_{10}^2|\varpi_{10}\rangle|\varpi_{10}\rangle\right)+|00\rangle_{12}|0\rangle_3|1\rangle_4\left(\delta_{00}\delta_{01}|\varpi_{00}\rangle|\varpi_{01}\rangle+\delta_{10}\delta_{11}|\varpi_{10}\rangle|\varpi_{11}\rangle\right)\\
&+|00\rangle_{12}|1\rangle_3|0\rangle_4\left(\delta_{01}\delta_{00}|\varpi_{01}\rangle|\varpi_{00}\rangle+\delta_{11}\delta_{10}|\varpi_{11}\rangle|\varpi_{10}\rangle\right)+|00\rangle_{12}|1\rangle_3|1\rangle_4\left(\delta_{01}^2|\varpi_{01}\rangle|\varpi_{01}\rangle+\delta_{11}^2|\varpi_{11}\rangle|\varpi_{11}\rangle\right)\\
&+|01\rangle_{12}|0\rangle_3|0\rangle_4\left(\delta_{10}\delta_{00}|\varpi_{10}\rangle|\varpi_{00}\rangle-\delta_{00}\delta_{10}|\varpi_{00}\rangle|\varpi_{10}\rangle\right)+|01\rangle_{12}|0\rangle_3|1\rangle_4\left(\delta_{10}\delta_{01}|\varpi_{10}\rangle|\varpi_{01}\rangle-\delta_{00}\delta_{11}|\varpi_{00}\rangle|\varpi_{11}\rangle\right)\\
&+|01\rangle_{12}|1\rangle_3|0\rangle_4\left(\delta_{11}\delta_{00}|\varpi_{11}\rangle|\varpi_{00}\rangle-\delta_{01}\delta_{10}|\varpi_{01}\rangle|\varpi_{10}\rangle\right)+|01\rangle_{12}|1\rangle_3|1\rangle_4\left(\delta_{11}\delta_{01}|\varpi_{11}\rangle|\varpi_{01}\rangle-\delta_{01}\delta_{11}|\varpi_{01}\rangle|\varpi_{11}\rangle\right)\\
&+|10\rangle_{12}|0\rangle_3|0\rangle_4\left(\delta_{00}\delta_{10}|\varpi_{00}\rangle|\varpi_{10}\rangle+\delta_{10}\delta_{00}|\varpi_{10}\rangle|\varpi_{00}\rangle\right)+|10\rangle_{12}|0\rangle_3|1\rangle_4\left(\delta_{00}\delta_{11}|\varpi_{00}\rangle|\varpi_{11}\rangle+\delta_{10}\delta_{01}|\varpi_{10}\rangle|\varpi_{01}\rangle\right)\\
&+|10\rangle_{12}|1\rangle_3|0\rangle_4\left(\delta_{01}\delta_{10}|\varpi_{01}\rangle|\varpi_{10}\rangle+\delta_{11}\delta_{00}|\varpi_{11}\rangle|\varpi_{00}\rangle\right)+|10\rangle_{12}|1\rangle_3|1\rangle_4\left(\delta_{01}\delta_{11}|\varpi_{01}\rangle|\varpi_{11}\rangle+\delta_{11}\delta_{01}|\varpi_{11}\rangle|\varpi_{01}\rangle\right)\\
&+|11\rangle_{12}|0\rangle_3|0\rangle_4\left(\delta_{00}^2|\varpi_{00}\rangle|\varpi_{00}\rangle-\delta_{10}^2|\varpi_{10}\rangle|\varpi_{10}\rangle\right)+|11\rangle_{12}|0\rangle_3|1\rangle_4\left(\delta_{00}\delta_{01}|\varpi_{00}\rangle|\varpi_{01}\rangle-\delta_{10}\delta_{11}|\varpi_{10}\rangle|\varpi_{11}\rangle\right)\\
&+|11\rangle_{12}|1\rangle_3|0\rangle_4\left(\delta_{01}\delta_{00}|\varpi_{01}\rangle|\varpi_{00}\rangle-\delta_{11}\delta_{10}|\varpi_{11}\rangle|\varpi_{10}\rangle\right)+|11\rangle_{12}|1\rangle_3|1\rangle_4\left(\delta_{01}^2|\varpi_{01}\rangle|\varpi_{01}\rangle-\delta_{11}^2|\varpi_{11}\rangle|\varpi_{11}\rangle\right)\Big]\\
=\frac{\sqrt{2}}{4}\Big(&|00\rangle_{12}|0\rangle_3|0\rangle_4|\Delta_{0000}\rangle+|00\rangle_{12}|0\rangle_3|1\rangle_4|\Delta_{0001}\rangle+|00\rangle_{12}|1\rangle_3|0\rangle_4|\Delta_{0010}\rangle+|00\rangle_{12}|1\rangle_3|1\rangle_4|\Delta_{0011}\rangle\\
&+|01\rangle_{12}|0\rangle_3|0\rangle_4|\Delta_{0100}\rangle+|01\rangle_{12}|0\rangle_3|1\rangle_4|\Delta_{0101}\rangle+|01\rangle_{12}|1\rangle_3|0\rangle_4|\Delta_{0110}\rangle+|01\rangle_{12}|1\rangle_3|1\rangle_4|\Delta_{0111}\rangle\\
&+|10\rangle_{12}|0\rangle_3|0\rangle_4|\Delta_{1000}\rangle+|10\rangle_{12}|0\rangle_3|1\rangle_4|\Delta_{1001}\rangle+|10\rangle_{12}|1\rangle_3|0\rangle_4|\Delta_{1010}\rangle+|10\rangle_{12}|1\rangle_3|1\rangle_4|\Delta_{1011}\rangle\\
&+|11\rangle_{12}|0\rangle_3|0\rangle_4|\Delta_{1100}\rangle+|11\rangle_{12}|0\rangle_3|1\rangle_4|\Delta_{1101}\rangle+|11\rangle_{12}|1\rangle_3|0\rangle_4|\Delta_{1110}\rangle+|11\rangle_{12}|1\rangle_3|1\rangle_4|\Delta_{1111}\rangle\Big), \quad (31)
\end{aligned}
$$

in accordance with Eq.(29) and Eq.(30), where $|\Delta_{0000}\rangle=\delta_{00}^2|\varpi_{00}\rangle|\varpi_{00}\rangle+\delta_{10}^2|\varpi_{10}\rangle|\varpi_{10}\rangle$, $|\Delta_{0001}\rangle=\delta_{00}\delta_{01}|\varpi_{00}\rangle|\varpi_{01}\rangle+\delta_{10}\delta_{11}|\varpi_{10}\rangle|\varpi_{11}\rangle$, $|\Delta_{0010}\rangle=\delta_{01}\delta_{00}|\varpi_{01}\rangle|\varpi_{00}\rangle+\delta_{11}\delta_{10}|\varpi_{11}\rangle|\varpi_{10}\rangle$, $|\Delta_{0011}\rangle=\delta_{01}^2|\varpi_{01}\rangle|\varpi_{01}\rangle+\delta_{11}^2|\varpi_{11}\rangle|\varpi_{11}\rangle$, $|\Delta_{0100}\rangle=\delta_{10}\delta_{00}|\varpi_{10}\rangle|\varpi_{00}\rangle-\delta_{00}\delta_{10}|\varpi_{00}\rangle|\varpi_{10}\rangle$, $|\Delta_{0101}\rangle=\delta_{10}\delta_{01}|\varpi_{10}\rangle|\varpi_{01}\rangle-\delta_{00}\delta_{11}|\varpi_{00}\rangle|\varpi_{11}\rangle$, $|\Delta_{0110}\rangle=\delta_{11}\delta_{00}|\varpi_{11}\rangle|\varpi_{00}\rangle-\delta_{01}\delta_{10}|\varpi_{01}\rangle|\varpi_{10}\rangle$, $|\Delta_{0111}\rangle=\delta_{11}\delta_{01}|\varpi_{11}\rangle|\varpi_{01}\rangle-\delta_{01}\delta_{11}|\varpi_{01}\rangle|\varpi_{11}\rangle$, $|\Delta_{1000}\rangle=\delta_{00}\delta_{10}|\varpi_{00}\rangle|\varpi_{10}\rangle+\delta_{10}\delta_{00}|\varpi_{10}\rangle|\varpi_{00}\rangle$, $|\Delta_{1001}\rangle=\delta_{00}\delta_{11}|\varpi_{00}\rangle|\varpi_{11}\rangle+\delta_{10}\delta_{01}|\varpi_{10}\rangle|\varpi_{01}\rangle$, $|\Delta_{1010}\rangle=\delta_{01}\delta_{10}|\varpi_{01}\rangle|\varpi_{10}\rangle+\delta_{11}\delta_{00}|\varpi_{11}\rangle|\varpi_{00}\rangle$, $|\Delta_{1011}\rangle=\delta_{01}\delta_{11}|\varpi_{01}\rangle|\varpi_{11}\rangle+\delta_{11}\delta_{01}|\varpi_{11}\rangle|\varpi_{01}\rangle$, $|\Delta_{1100}\rangle=\delta_{00}^2|\varpi_{00}\rangle|\varpi_{00}\rangle-\delta_{10}^2|\varpi_{10}\rangle|\varpi_{10}\rangle$, $|\Delta_{1101}\rangle=\delta_{00}\delta_{01}|\varpi_{00}\rangle|\varpi_{01}\rangle-\delta_{10}\delta_{11}|\varpi_{10}\rangle|\varpi_{11}\rangle$, $|\Delta_{1110}\rangle=\delta_{01}\delta_{00}|\varpi_{01}\rangle|\varpi_{00}\rangle-\delta_{11}\delta_{10}|\varpi_{11}\rangle|\varpi_{10}\rangle$ and $|\Delta_{1111}\rangle=\delta_{01}^2|\varpi_{01}\rangle|\varpi_{01}\rangle-\delta_{11}^2|\varpi_{11}\rangle|\varpi_{11}\rangle$.

(i) Consider the case that Alice chooses the MEASURE_A2 mode and Bob chooses the REFLECT_B mode. In order to escape the security check in Case ③ of Step 5, Eve should make

TP's Bell basis measurement result on the corresponding particles of $S_1$ and $S_2$ correctly correlate to Alice's $Z$ basis measurement result on the particle pair of $S_3$, in accordance with Eq.(7). Hence, it should satisfy that

$$|\Delta_{0100}\rangle = |\Delta_{1000}\rangle = |\Delta_{0111}\rangle = |\Delta_{1011}\rangle = |\Delta_{0001}\rangle = |\Delta_{1101}\rangle = |\Delta_{0010}\rangle = |\Delta_{1110}\rangle = 0, \quad (32)$$

$$|\Delta_{0000}\rangle = |\Delta_{1100}\rangle, \quad (33)$$

$$|\Delta_{0011}\rangle = -|\Delta_{1111}\rangle, \quad (34)$$

$$|\Delta_{0101}\rangle = -|\Delta_{1001}\rangle \quad (35)$$

and

$$|\Delta_{0110}\rangle = |\Delta_{1010}\rangle. \quad (36)$$

Inserting Eqs.(32-36) into Eq.(31) has

$$\begin{aligned} U_E\left(|\chi^{00}\rangle_{1234}|\Omega\rangle\right) = \frac{\sqrt{2}}{4}\big(&|00\rangle_{12}|0\rangle_3|0\rangle_4|\Delta_{0000}\rangle + |00\rangle_{12}|1\rangle_3|1\rangle_4|\Delta_{0011}\rangle + |01\rangle_{12}|0\rangle_3|1\rangle_4|\Delta_{0101}\rangle + |01\rangle_{12}|1\rangle_3|0\rangle_4|\Delta_{0110}\rangle \\ &+ |10\rangle_{12}|0\rangle_3|1\rangle_4|\Delta_{1001}\rangle + |10\rangle_{12}|1\rangle_3|0\rangle_4|\Delta_{1010}\rangle + |11\rangle_{12}|0\rangle_3|0\rangle_4|\Delta_{1100}\rangle + |11\rangle_{12}|1\rangle_3|1\rangle_4|\Delta_{1111}\rangle\big) \\ = \frac{1}{2}\big(&|\phi^+\rangle_{12}|0\rangle_3|0\rangle_4|\Delta_{0000}\rangle + |\phi^-\rangle_{12}|1\rangle_3|1\rangle_4|\Delta_{0011}\rangle - |\psi^-\rangle_{12}|0\rangle_3|1\rangle_4|\Delta_{1001}\rangle + |\psi^+\rangle_{12}|1\rangle_3|0\rangle_4|\Delta_{0110}\rangle\big). \quad (37) \end{aligned}$$

(ii) Consider the case that Alice chooses the MEASURE_A2 mode and Bob chooses the MEASURE_B2 mode. According to Eq.(37), after Alice performs the MEASURE_A2 mode, the global state of the composite system is randomly collapsed into $|\phi^+\rangle_{12}|0\rangle_3|0\rangle_4|\Delta_{0000}\rangle$, $|\phi^-\rangle_{12}|1\rangle_3|1\rangle_4|\Delta_{0011}\rangle$, $|\psi^-\rangle_{12}|0\rangle_3|1\rangle_4|\Delta_{1001}\rangle$ or $|\psi^+\rangle_{12}|1\rangle_3|0\rangle_4|\Delta_{0110}\rangle$. For Eve not being detected in Case ④ of Step 5, Alice's measurement result on the particle of $S_3$ should be same to Bob's measurement result on the corresponding particle of $S_3^{'}$. As a result, the whole quantum system after being applied with $U_F$ should be

$$U_F\left(|\phi^+\rangle_{12}|0\rangle_3|0\rangle_4|\Delta_{0000}\rangle\right) = |\phi^+\rangle_{12}|0\rangle_3|0\rangle_4|F_{0000}\rangle, \quad (38)$$

$$U_F\left(|\phi^-\rangle_{12}|1\rangle_3|1\rangle_4|\Delta_{0011}\rangle\right) = |\phi^-\rangle_{12}|1\rangle_3|1\rangle_4|F_{0011}\rangle, \quad (39)$$

$$U_F\left(|\psi^-\rangle_{12}|0\rangle_3|1\rangle_4|\Delta_{1001}\rangle\right) = |\psi^-\rangle_{12}|0\rangle_3|1\rangle_4|F_{1001}\rangle \quad (40)$$

or

$$U_F\left(|\psi^+\rangle_{12}|1\rangle_3|0\rangle_4|\Delta_{0110}\rangle\right) = |\psi^+\rangle_{12}|1\rangle_3|0\rangle_4|F_{0110}\rangle, \quad (41)$$

which means that $U_F$ cannot alter the state of the fresh particle in $S_3^{'}$ prepared by Alice after her measurement on the particle of $S_3$.

(iii) Consider the case that Alice chooses the REFLECT_A mode and Bob chooses the MEASURE_B2 mode. On the basis of Eq.(37) and Eqs.(38-41), the whole quantum system after being applied with $U_F$ should be

$$
\begin{aligned}
&U_F\left[U_E\left(\left|\chi^{00}\right\rangle_{1234}|\Omega\rangle\right)\right] \\
&=U_F\left[\frac{1}{2}\left(\left|\phi^{+}\right\rangle_{12}|0\rangle_3|0\rangle_4|\Delta_{0000}\rangle+\left|\phi^{-}\right\rangle_{12}|1\rangle_3|1\rangle_4|\Delta_{0011}\rangle-\left|\psi^{-}\right\rangle_{12}|0\rangle_3|1\rangle_4|\Delta_{1001}\rangle+\left|\psi^{+}\right\rangle_{12}|1\rangle_3|0\rangle_4|\Delta_{0110}\rangle\right)\right] \\
&=\frac{1}{2}\left(\left|\phi^{+}\right\rangle_{12}|0\rangle_3|0\rangle_4|F_{0000}\rangle+\left|\phi^{-}\right\rangle_{12}|1\rangle_3|1\rangle_4|F_{0011}\rangle-\left|\psi^{-}\right\rangle_{12}|0\rangle_3|1\rangle_4|F_{1001}\rangle+\left|\psi^{+}\right\rangle_{12}|1\rangle_3|0\rangle_4|F_{0110}\rangle\right). \quad (42)
\end{aligned}
$$

After Bob implements the MEASURE_B2 mode, the whole quantum system is randomly collapsed into $\left|\phi^{+}\right\rangle_{12}|0\rangle_3|0\rangle_4|F_{0000}\rangle$, $\left|\phi^{-}\right\rangle_{12}|1\rangle_3|1\rangle_4|F_{0011}\rangle$, $\left|\psi^{-}\right\rangle_{12}|0\rangle_3|1\rangle_4|F_{1001}\rangle$ or $\left|\psi^{+}\right\rangle_{12}|1\rangle_3|0\rangle_4|F_{0110}\rangle$. As a result, the presence of Eve cannot be discovered in Case ② of Step 5.

(iv) Consider the case that Alice chooses the REFLECT_A mode and Bob chooses the REFLECT_B mode. For Eve not being discovered in Case ① of Step 5, TP's *FMB* basis measurement result on the particle pair of $S_3^{''}$, the corresponding particle of $S_1$ and the corresponding particle of $S_2$ should always be $\left|\chi^{00}\right\rangle$. Thus, by virtue of Eq.(42), the whole quantum system after the operation of Bob should be

$$
\begin{aligned}
&U_F\left[U_E\left(\left|\chi^{00}\right\rangle_{1234}|\Omega\rangle\right)\right] \\
&=\frac{1}{2}\left(\left|\phi^{+}\right\rangle_{12}|0\rangle_3|0\rangle_4|F_{0000}\rangle+\left|\phi^{-}\right\rangle_{12}|1\rangle_3|1\rangle_4|F_{0011}\rangle-\left|\psi^{-}\right\rangle_{12}|0\rangle_3|1\rangle_4|F_{1001}\rangle+\left|\psi^{+}\right\rangle_{12}|1\rangle_3|0\rangle_4|F_{0110}\rangle\right) \\
&=\frac{1}{2}\left(\left|\phi^{+}\right\rangle_{12}|0\rangle_3|0\rangle_4+\left|\phi^{-}\right\rangle_{12}|1\rangle_3|1\rangle_4-\left|\psi^{-}\right\rangle_{12}|0\rangle_3|1\rangle_4+\left|\psi^{+}\right\rangle_{12}|1\rangle_3|0\rangle_4\right)\left|F^{'}\right\rangle \\
&=\left|\chi^{00}\right\rangle_{1234}\left|F^{'}\right\rangle. \quad (43)
\end{aligned}
$$

We can derive from Eq.(43) that

$$|F_{0000}\rangle=|F_{0011}\rangle=|F_{1001}\rangle=|F_{0110}\rangle=\left|F^{'}\right\rangle. \quad (44)$$

(v) Applying Eq.(44) into Eqs.(38-41) produces

$$U_F\left(\left|\phi^{+}\right\rangle_{12}|0\rangle_3|0\rangle_4|\Delta_{0000}\rangle\right)=\left|\phi^{+}\right\rangle_{12}|0\rangle_3|0\rangle_4\left|F^{'}\right\rangle, \quad (45)$$

$$U_F\left(\left|\phi^{-}\right\rangle_{12}|1\rangle_3|1\rangle_4|\Delta_{0011}\rangle\right)=\left|\phi^{-}\right\rangle_{12}|1\rangle_3|1\rangle_4\left|F^{'}\right\rangle, \quad (46)$$

$$U_F\left(\left|\psi^{-}\right\rangle_{12}|0\rangle_3|1\rangle_4|\Delta_{1001}\rangle\right)=\left|\psi^{-}\right\rangle_{12}|0\rangle_3|1\rangle_4\left|F^{'}\right\rangle \quad (47)$$

and

$$U_F\left(\left|\psi^+\right\rangle_{12}\left|1\right\rangle_3\left|0\right\rangle_4\left|\Delta_{0110}\right\rangle\right)=\left|\psi^+\right\rangle_{12}\left|1\right\rangle_3\left|0\right\rangle_4\left|F'\right\rangle \quad (48)$$

It can be concluded from that Eq.(43) and Eqs.(45-48), when Eve imposes $U_E$ on the particle pair of $S_3$ from TP to Alice and $U_F$ on the particle pair of $S_3^{'}$ from Alice to Bob in Step 4, for introducing no error in Step 5, the final state of Eve's probe should be independent of not only the operations of Alice and Bob but also their measurement results. As a result, Eve acquires no information about $K$ .

(4) The Trojan horse attacks

Since the particles of $S_3$ go a round trip among TP, Alice, Bob and TP, we also need to take actions to prevent the Trojan horse attacks. Similar to Sect.4.1.1, the receiver can utilize a wavelength filter to defeat the invisible photon eavesdropping attack and a photon number splitter (PNS: 50/50) to overcome the delay-photon Trojan horse attack [34,35].

### 4.2 Participant attacks

In 2007, Gao *et al.* [36] put forward for the first time that participant attacks must be paid special concerns to, since participants always have strong powers than an outside attacker. In the following, we will analyze three types of participant attacks in detail.

(1) The participant attack from dishonest Alice

Suppose that Alice is a dishonest semiquantum user who attempts to steal $p_b$ . If Alice announces the wrong operation modes for the received particles of $S_1$ in Step 3, due to being unknown about the corresponding operation modes Bob chose, her cheating behaviors will be inevitably detected by TP and Bob during the security check processes of Step 3. As a result, Alice has to honestly announce her operation modes for the received particles of $S_1$ in Step 3. Hence, Alice has no chance to obtain $m_b$ , which is derived from the Case that Alice has entered into the REFLECT_A mode, while Bob has entered into the MEASURE_B mode. Although Alice knows $K$ and may hear of $f$ sent from Bob to TP, she still has no chance to obtain $p_b$ , which is encrypted with $m_b$ , according to Eq.(10).

In addition, although Alice receives the final comparison result between $p_a$ and $p_b$ from TP in Step 7, Alice still has no way to obtain $p_b$ .

(2) The participant attack from dishonest Bob

Suppose that dishonest semiquantum user Bob wants to extract $p_a$ . If Bob announces the wrong operation modes for the received particles of $S_1^{'}$ in Step 3, his cheating behaviors will be undoubtedly discovered by TP and Alice during the security check processes of Step 3, because of being short of the corresponding operation modes Alice chose. Hence, Bob has to announce the correct operation modes for the received particles of $S_1^{'}$ in Step 3. In this way, Bob is unknown about $m_a$, which is derived from the Case that Alice has entered into the MEASURE_A mode, while Bob has entered into the REFLECT_B mode. Although Bob gets $K$ and may hear of $g$ sent from Alice to TP, he still cannot decode out $p_a$ from $g$, as $p_a$ is encrypted by $m_a$, according to Eq.(9).

In addition, although Bob hears of the final comparison result between $p_a$ and $p_b$ from TP in Step 7, Bob still has no chance to know $p_a$.

(3) The participant attack from semi-honest TP

In the proposed SQPC protocol, TP is permitted to misbehave on her own but cannot collude with anyone else [12]. Obviously, TP is able to obtain $m_a$ and $m_b$ easily. According to Eq.(9) and Eq.(10), in order to steal $p_a$ and $p_b$, TP needs to acquire $K$ first. Apparently, $K$ is derived from the $L$ bits of $M_A$ used for generated the fresh quantum states by Alice in Case ④, where Alice has entered into the MEASURE_A2 mode, and Bob has entered into the MEASURE_B2 mode. TP has no opportunity to obtain the accurate values of $K$ due to the following two reasons: on the one hand, $M_A$ is random to TP; and on the other hand, $M_B$ is irrelevant to $M_A$. As a result, TP cannot obtain $p_a$ and $p_b$.

Furthermore, although TP gets the final comparison result between $p_a$ and $p_b$ in Step 7, it is still helpless for her to know the accurate values of $p_a$ and $p_b$.

## 5 Discussions

In the following, we will calculate the qubit efficiency for the proposed SQPC protocol, which is defined as [4]

$$\eta = \frac{\alpha}{\beta + \gamma}, \tag{49}$$

where $\alpha$, $\beta$ and $\gamma$ are the length of compared private binary string, the number of consumed qubits and the number of classical bits consumed during the classical communication, respectively. Note that the classical resources used for eavesdropping detections are ignored here.

In the proposed SQPC protocol, the length of $p_a$ or $p_b$ is $L$ bits, so we get $\alpha = L$. TP generates $12L$ χ-type states and splits them into $S_1$, $S_2$ and $S_3$. Then TP sends the first $8L$ particles in $S_1$ to Alice one by one; when Alice chooses the MEASURE_A mode, she uses the $Z$ basis to measure the received particles, writes down the measurement results, produces the fresh quantum states as found and transmits them to Bob; when Bob chooses the MEASURE_B mode, he adopts the $Z$ basis to measure the received particles, writes down the measurement results, generates the fresh quantum states as found and returns them to TP. Afterward, TP sends the last $4L$ particle pairs in $S_3$ to Alice; when Alice chooses the MEASURE_A2 mode, she uses the $Z$ basis to measure the received particles, writes down the measurement results, generates the fresh quantum states according to the found states or the bits of $M_A$ with equal probabilities and sends them to Bob; when Bob chooses the MEASURE_B2 mode, he employs the $Z$ basis to measure the received particles, writes down the measurement results, produces the fresh quantum states according to the bits of $M_B$ and returns them to TP. Hence, it has $\beta = 12L \times 4 + 4L \times 2 + 2L \times 2 \times 2 = 64L$. Furthermore, Alice transmits $g$ to TP, while Bob sends $f$ to TP, so we obtain $\gamma = L + L = 2L$. It can be concluded that the qubit efficiency of the proposed SQPC protocol is $\eta = \frac{L}{64L + 2L} = \frac{1}{66}$.

In addition, we compare the proposed SQPC protocol with previous SQPC protocols based on quantum entangled states. The specific comparison outcomes are listed in Table 3.

In the SQPC protocol of Ref.[22], the length of Alice's or Bob's private binary sequence is $n$ bits, so it has $\alpha = n$. TP produces two Bell state sequences of length $16n$, and sends the first particles of two sequences to Alice and Bob, respectively; when Alice and Bob decide to measure the received qubits, they generate the fresh qubits to substitute the measurement results. Hence, we obtain $\beta = 16n \times 2 \times 2 + 8n \times 2 = 80n$. Moreover, Alice sends $R_A$ to TP; in the meanwhile, Bob sends $R_B$ to TP. As a result, we get $\gamma = 2n$. It can be concluded that the qubit efficiency of the protocol in Ref.[22] is $\eta = \frac{n}{80n + 2n} = \frac{1}{82}$.

In the SQPC protocol of Ref.[23], the length of Alice's or Bob's private binary sequence is $n$ bits, so it has $\alpha = n$. TP generates $8n$ Bell states, and sends the first particles to Alice and the second particles to Bob; when Alice and Bob choose the MEASURE mode, they generate the fresh qubits to substitute the measurement results. Moreover, this protocol utilizes the SQKD protocol of Ref.[37] to establish $K_{AB}$, consuming $24n$ qubits, and the SQKA protocol of Ref.[38] to produce $K_{AT}$ and $K_{BT}$, totally consuming $10n$ qubits. Hence, we obtain $\beta = 8n \times 2 + 4n \times 2 + 24n + 10n = 58n$. In addition, Alice transmits $C_A$ to TP; in the meanwhile, Bob transmits $C_B$ to TP. As a result, we obtain $\gamma = 2n$. It can be concluded that the qubit efficiency of the protocol in Ref.[23] is $\eta = \frac{n}{58n + 2n} = \frac{1}{60}$.

In the SQPC protocol of Ref.[26], the length of the hash values of Alice's or Bob's private binary sequence is $n$ bits, so we get $\alpha = n$. Server generates $4n$ Bell states, and sends the first particles to Alice and the second particles to Bob; and then, Alice and Bob write down the measurement results or return the received qubits after reordering them. Hence, we obtain $\beta = 4n \times 2 = 8n$. Furthermore, Alice transmits $R_A$ to Bob; in the meanwhile, Bob transmits $R_B$ to Alice. As a result, we get $c = 2n$. It can be concluded that the qubit efficiency of the protocol in Ref.[23] is $\eta = \frac{n}{8n + 2n} = \frac{1}{10}$. Note that Server gets the comparison result by calculating $R_A \oplus R_B$ after $R_A$ and $R_B$ are announced by Alice and Bob, respectively.

In the second SQPC protocol of Ref.[27], the length of Alice's or Bob's private binary sequence is $n$ bits, hence we get $\alpha = n$. TP generates $2n$ Bell states, and sends the first particles to Alice and the second particles to Bob; when Alice and Bob choose to flip the received qubits, they use the freshly opposite qubits to substitute the measurement results. Moreover, this protocol uses the SQKD protocol of Ref.[37] to produce $K$, consuming $24n$ qubits. Hence, we obtain $\beta = 2n\times2+n\times2+24n=30n$. In addition, Alice transmits $M_A$ to TP; in the meanwhile, Bob transmits $M_B$ to TP. As a result, we get $\gamma = 2n$. It can be concluded that the qubit efficiency of the second protocol of Ref.[27] is $\eta = \frac{n}{30n+2n} = \frac{1}{32}$.

In the SQPC protocol of Ref.[28], the length of Alice's or Bob's private binary sequence is $n$ bits, hence we get $\alpha = n$. TP produces $2n$ Bell states, and sends the first particles to Alice and the second particles to Bob; when Alice and Bob select the SIFT operation, they send new qubits to TP after preparing them. Moreover, this protocol adopts the SQKD protocol of Ref.[18] to prepare $K_{TA}$ and $K_{TB}$, totally consuming $16n$ qubits, and the SQKD protocol of Ref.[37] to produce $K$, consuming $24n$ qubits. Hence, we get $\beta = 2n\times2+n\times2+16n+24n=46n$. In addition, Alice sends $R_A$ to TP; in the meanwhile, Bob sends $R_B$ to TP. As a result, we obtain $\gamma = 2n$. It can be concluded that the qubit efficiency of the protocol in Ref.[28] is $\eta = \frac{n}{46n+2n} = \frac{1}{48}$.

In the SPQC protocol of Ref.[29], the length of Alice's or Bob's private binary sequence is $n$ bits, hence we get $\alpha = n$. TP generates $3n$ Bell states, and sends the first particles to Alice and the second particles to Bob; when Alice and Bob select the SIFT operation, they send new qubits to TP after generating them; and when Alice and Bob select the DETECT operation, they send new trap qubits to TP after producing them. Moreover, this protocol employs the SQKD protocol of Ref.[37] to produce $K$, consuming $24n$ qubits. Hence, we obtain $\beta = 3n\times2+n\times2+n\times2+24n=34n$. In addition, Alice sends $R_A \oplus R_A^{'}$ to TP; in the meanwhile, Bob sends $R_B \oplus R_B^{'}$ to TP. As a result, we obtain $\gamma = 2n$. It can be concluded that the qubit efficiency of the protocol in Ref.[29] is $\eta = \frac{n}{34n+2n} = \frac{1}{36}$.

In the SQPC protocol of Ref.[31], the length of Alice's or Bob's private binary sequence is $n$ bits, hence we get $\alpha = n$. TP produces $2n$ Bell states, and sends the first particles to Alice and the second particles to Bob; when Alice and Bob select the MEASURE mode, they send new qubits to TP after preparing them. Furthermore, this protocol adopts the SQKD protocol of Ref.[37] to produce $K_{AB}$, consuming $48n$ qubits. Hence, we obtain $\beta = 2n\times2+n\times2+48n=54n$. In addition, Alice sends $C_A$ to TP; in the meanwhile, Bob sends $C_B$ to TP. Consequently, it has $\gamma = 4n$. It can be concluded that the qubit efficiency of the protocol in Ref.[31] is $\eta = \frac{n}{54n+4n} = \frac{1}{58}$.

In the SQPC protocol of Ref.[32], the length of Alice's or Bob's private binary sequence is $n$ bits, hence we get $\alpha = n$. TP produces $4n$ W states, and sends the first particles to Alice and the second particles to Bob; when Alice and Bob select the SIFT mode, they measure the received particles with the $Z$ basis and send TP new qubits identical to their measurement results.

Furthermore, this protocol employs the SQKD protocol of Ref.[37] to produce $K_{AB}$, consuming $24n$ qubits. Hence, it has $\beta = 4n \times 3 + 2n \times 2 + 24n = 40n$. In addition, both Alice and Bob send $K_{AB}$ to TP. As a result, we obtain $\gamma = 2n$. It can be concluded that the qubit efficiency of the protocol in Ref.[32] is $\eta = \frac{n}{40n + 2n} = \frac{1}{42}$.

In the SQPC protocol of Ref.[33], the length of Alice's or Bob's private binary sequence is $n$ bits, hence we get $\alpha = n$. TP generates $N = 16n$ Bell states, sends both $T_1$ and $T_5$ to Alice, and transmits both $T_3$ and $T_6$ to Bob; when Alice selects the MEASURE mode for the received qubits of $T_1$, she generates $2n$ fresh qubits; when Alice selects the MEASURE mode for the received qubits of $T_5$, she produces $4n$ fresh qubits; when Bob chooses the MEASURE mode for the received qubits in $T_3$, he produces $2n$ fresh qubits; when Bob chooses the MEASURE mode for the received qubits in $T_6$, he generates $4n$ fresh qubits. Furthermore, this protocol utilizes the SQKD protocol of Ref.[37] to produce $K_{AB}$, consuming $24n$ qubits. Hence, we obtain $\beta = 16n \times 2 + 2n \times 2 + 4n \times 2 + 24n = 68n$. In addition, Alice sends $R_A$ to TP; in the meanwhile, Bob sends $R_B$ to TP. As a result, we obtain $\gamma = 2n$. It can be concluded that the qubit efficiency of the protocol in Ref.[33] is $\eta = \frac{n}{68n + 2n} = \frac{1}{70}$.

In accordance with Table 3, we can know that the proposed SQPC protocol exceeds the protocol of Ref.[22] on the usage of quantum entanglement swapping, as it has no demand for quantum entanglement swapping; on the usage of a pre-shared key, the proposed SQPC protocol is superior to the protocols of Refs.[23,27-29,31-33], as it doesn't require a pre-shared key; on the usage of delay lines, the proposed SQPC protocol is superior to the protocols of Refs.[23,26,27], as it doesn't use any delay lines; as for the qubit efficiency, the proposed SQPC protocol exceeds the protocols of Refs.[22,33].

## 6 Conclusions

In this paper, a novel circular SQPC protocol with $\chi$-type states without a pre-shared key is put forward, which accomplishes the equality comparison of private inputs from two different semiquantum users under the help of a quantum TP within one execution of protocol. The quantum TP can try her best to launch all kinds of attacks but cannot be allowed to collude with others. The travelling qubits go a round trip among TP, Alice and Bob. We have proven in detail that the proposed SQPC protocol is immune to both the outside attacks and the inside attacks. The proposed SQPC protocol has no demand for unitary operations. The proposed SQPC protocol exceeds the previous SQPC protocols of equality based on quantum entangled states in Refs.[22,23,26-29,31-33] in the following four aspects more or less: ① it requires no pre-shared key among different participants; ② it doesn't need quantum entanglement swapping; and ③ it employs no delay lines.

Table 3  Comparison results between the proposed SQPC protocol and previous SQPC protocols

| | Feature | Comparison of equality | Transmission mode | Quantum resources | Type of TP | Usage of quantum entanglement swapping | Usage of unitary operations | Usage of pre-shared key | Usage of delay lines | TP's knowledge about the comparison result | Qubit efficiency |
|---|---|---|---|---|---|---|---|---|---|---|---|
| The protocol of Ref.[22] | Measure-resend | Yes | Tree-type | Bell states | Semi-honest | Yes | No | No | No | Yes | $\frac{1}{82}$ |
| The protocol of Ref.[23] | Measure-resend | Yes | Tree-type | Bell states | Semi-honest | No | No | Yes | Yes | Yes | $\frac{1}{60}$ |
| The protocol of Ref.[26] | Randomization-based | Yes | Tree-type | Bell states | Semi-honest | No | No | No | Yes | Yes | $\frac{1}{10}$ |
| The second protocol of Ref.[27] | Measure-Randomization-resend | Yes | Tree-type | Bell states | Semi-honest | No | No | Yes | Yes | Yes | $\frac{1}{32}$ |
| The protocol of Ref.[28] | Discard-resend | Yes | Tree-type | Bell states | Semi-honest | No | No | Yes | No | Yes | $\frac{1}{48}$ |
| The protocol of Ref.[29] | Measure-discard-resend | Yes | Tree-type | Bell states | Semi-honest | No | No | Yes | No | Yes | $\frac{1}{36}$ |
| The protocol of Ref.[31] | Measure-resend | Yes | Tree-type | Bell states | Semi-honest | No | No | Yes | No | Yes | $\frac{1}{58}$ |
| The protocol of Ref.[32] | Measure-resend | Yes | Tree-type | W states | Semi-honest | No | No | Yes | No | Yes | $\frac{1}{42}$ |
| The protocol of Ref.[33] | Measure-resend | Yes | Tree-type | Bell states | Semi-honest | No | No | Yes | No | Yes | $\frac{1}{70}$ |
| The proposed protocol | Measure -resend | Yes | Circular | $\chi$-type states | Semi-honest | No | No | No | No | Yes | $\frac{1}{66}$ |

## Acknowledgments

Funding by the National Natural Science Foundation of China (Grant No.62071430), the Fundamental Research Funds for the Provincial Universities of Zhejiang (Grant No. JRK21002) and the General Project of Zhejiang Provincial Education Department (Grant No. Y202250189) is gratefully acknowledged.